\documentclass[reprint,scriptaddress,prd,showkeys,11pt]{article}
 \usepackage{jcappub} 

	\usepackage{amsmath}
	\usepackage{makeidx}
	\usepackage{amsfonts}
	\usepackage[ansinew]{inputenc}
	\usepackage[usenames,dvipsnames]{pstricks}
	\usepackage{epsfig}
	\usepackage{color}
	\usepackage{pst-grad} 
	\usepackage{pst-plot} 
\usepackage{booktabs}

	\newcommand{\submin}[1]{\left\langle #1 \right\rangle}

	\newcommand{\argu}[1]{\left( #1 \right)}
	
	\newcommand{\beq}{\begin{equation}}
	\newcommand{\ti}[1]{\tilde{#1}}
\newcommand{\eeq}{\end{equation}}
\newcommand{\bea}{\begin{eqnarray}}
\newcommand{\eea}{\end{eqnarray}}
\usepackage{color}
\usepackage{ulem}

\makeindex

\begin{document}

\title{Cosmological Perturbations in the 5D Holographic Big Bang Model}

\author[a,b]{Natacha Altamirano}
\author[a,b]{Elizabeth Gould}
\author[a,b]{Niayesh Afshordi}
\author[b]{Robert B. Mann}


\affiliation[a]{Perimeter Institute for Theoretical Physics,\\31 Caroline St. N., Waterloo, ON, N2L 2Y5, Canada}
\affiliation[b]{Department of Physics and Astronomy, University of Waterloo,\\ Waterloo, ON, N2L 3G1, Canada}

\emailAdd{naltamirano@pitp.ca}
\emailAdd{egould@pitp.ca}
\emailAdd{nafshordi@pitp.ca}
\emailAdd{rbmann@uwaterloo.ca}
\date{\today}

\abstract{
The 5D Holographic Big Bang is a novel model for the emergence of the early universe out of a 5D collapsing star (an apparent white hole),
in the context of  Dvali-Gabadadze-Porrati (DGP) cosmology. The model does not have a big bang singularity, and yet can
address cosmological puzzles that are traditionally solved within inflationary cosmology. 
In this paper, we compute the exact power spectrum of  cosmological  curvature perturbations due to the effect
of a thin atmosphere accreting into our 3-brane. The spectrum is scale-invariant on small scales and red on intermediate scales, but becomes blue on scales larger than the height of the atmosphere. While this behaviour  is broadly consistent with the non-parametric measurements of the primordial scalar power spectrum, it is marginally disfavoured relative to a simple power law (at  2.7$\sigma$ level). Furthermore, we find that the best fit nucleation temperature of our 3-brane is  at least  3 orders of magnitude larger than the 5D Planck mass, suggesting an origin in a 5D quantum gravity phase. }

\keywords{Cosmology, Early Universe, DGP, Scalar Power Spectrum}

\maketitle


\section{Introduction}

Modern cosmology continues to experience an astonishing degree of empirical 
success \cite{Plankresults2015}. The agreement with the phenomenological six-parameter  $\Lambda$CDM paradigm 
is remarkable, all the more so as the number of
 cosmological observations  continue to increase at an accelerated rate.
 
Despite this success, there are still intriguing puzzles
left unresolved: the big bang singularity, the horizon and
flatness problems (traditionally addressed within the inflationary paradigm), as well as the nature of dark matter and dark energy. Recently \cite{Holobigbang2013}, a novel cosmological model was proposed in which our 
universe is a 3-brane emergent from the collapse of a 5-dimensional
star.  Motivated by the desire to see if a more 
satisfactory (or natural) understanding of these puzzles can
emerge from an alternative description of the geometry, this
 model explains the evolution of our early universe whilst avoiding a big bang singularity. 
 Furthermore, the model was shown to have a
  mechanism via which a homogeneous  atmosphere outside the black hole generates a scale invariant 
power spectrum for primordial curvature perturbations, (nearly) consistent with current cosmological observations \cite{Plankresults2015}. 

Our 5D holographic origin for the big bang is based on a braneworld 
theory that includes both  4 dimensional induced gravity {\it and}  5D bulk gravity: 
the Dvali-Gabadadze-Porrati (DGP) model \cite{Dvali:2000hr},
with action
\beq
S_{DGP}=\frac{1}{16\pi G_5}\int_{\text{bulk}}\!\!\!d^5x\sqrt{-g}R_5+\frac{1}{8\pi G_5}
\int_{\text{brane}}\!\!\!d^4x\sqrt{-\gamma}K+\int_{\text{brane}}\!\!\!\sqrt{-\gamma}\bigg(\frac{R_4}{16\pi G_4}+
{\cal{L}}_{\text{matter}}\bigg)\,,
\label{DGPaction}
\eeq
\\
 where $g$ and $\gamma$, $G_5$ and $G_4$, $R_5$ and $R_4$ are the metrics, gravitational constants and Ricci scalars of 
 the bulk and brane respectively, while $K$ is the mean extrinsic curvature
 of the brane. Our universe, described by the metric
  \beq
 ds^2_4=-d\tau^2+\frac{a^2(\tau)}{{\cal{K}}}[d\psi^2+\sin^2(\psi)(d\theta^2+\sin^2(\theta)d\phi^2)]\,,
 \eeq
 is represented by a hypersurface in a 5 dimensional Schwarzschild black hole
 spacetime
  \beq
 ds^2_5=-\left(1-\frac{\mu}{r^2}\right)dt^2+\left(1-\frac{\mu}{r^2}\right)^{-1} dr^2+r^2d\Omega_3^2~,
 \label{bulk_sch}
 \eeq
 located at $r=\frac{a(\tau)}{\sqrt{{\cal{K}}}}$. In this context, a pressure singularity is generically found when the energy density of the holographic fluid $\ti{\rho}$ satisfies $\ti{\rho}=\ti{\rho}_s=\frac{3G_4}{16\pi G_5^2}$
 \cite{Gregory:2007xy}. The authors in \cite{Holobigbang2013} showed that the singularity happens
 at early times in the cosmic history as matter decays more slowly than $a^{-4}$. 
 However, under the evolution from smooth initial conditions, the pressure singularity
 can occur before Big Bang Nucleosynthesis (BBN), and is generically inside a white hole horizon. Alternatively, the universe  could have emerged from the collapse of
 a 5D star into a black hole, just before BBN, removing both pressure {\it and} big bang/white hole singularities.   As advocated in the {\it fuzzball} program \cite{Mathur:2008kg}, the rate of this tunnelling is enhanced due to the large entropy of black hole microstates, which we speculate could match those of an expanding 3-brane thermal state. Our universe is represented by the boundary of a 5D spherically symmetric spacetieme with metric \eqref{bulk_sch}, 
in which we impose ${\mathbb{Z}}_2$ boundary conditions. This picture will be described in more detail in Section \ref{strategy}.
 
 Interestingly, this model not only  circumvents the singularity at the origin of time, but  can also address other problems of cosmology that are typically  solved by inflation. Because  the collapsing star could have existed long before its demise,
 it had enough time to attain uniform temperature, thereby addressing the {\it{Horizon Problem}}. 
 Furthermore, if we assume that the initial Hubble constant was of order of the 5D Planck mass, then the curvature density
 $-\Omega_k\sim(M_5r_h)^{-2}$, where $r_h$ is the radius of the black hole.  Consequently $-\Omega_k
 \sim M_5/M_{*}$ could become very small for massive stars, thus solving the {\it{Flatness Problem}}. More generically, the {\it no hair} theorem ensures that a 3-brane nucleated just outside the event horizon of a massive black hole has a smooth geometry.
 
 Yet another feature of the model is that a thermal atmosphere 
 around the brane, composed of a gas of massless particles
 produces scale invariant curvature perturbations. 
 In this work, we shall revisit this result, focusing our attention on the mechanism responsible for deviations from scale-invariance  in the primordial curvature power spectrum in the context of the
 {\it{5D Holographic origin of the Big Bang}}. To this end, we consider a thin atmosphere
 that can be regarded as infalling matter, or the 
 outer envelope of the collapsing 5D star, which resides in the 5D bulk and thus contributes to its energy
 momentum tensor.  In this context, the 
DGP action Eq. \eqref{DGPaction} is modified to 
\beq
S= S_{DGP}+\int_{\text{bulk}}d^5x\sqrt{-g}{\cal{L}}_{\text{5,atmosphere}}\label{DGPaction1}
\eeq
where ${\cal{L}}_{\text{5,atmosphere}}$ accounts for the matter Lagrangian in the bulk. Consistency
between cosmological phenomenology and the DGP model implies that the brane is expanding outwards, and thus eventually encounters this {\it{atmosphere}}. We then compute the resulting power spectrum of scalar curvature perturbations and the change of the Hubble parameter due to this encounter.
 We find that the best fit nucleation temperature of the 3-brane is considerably larger than the 5D Planck mass, perhaps indicating an origin in a 5D quantum gravity phase. 

In 
Section \ref{strategy}, we  discuss a possible mechanism for brane nucleation and 
the different scales involved in 
our problem, giving a qualitative description of the different physics processes.  
In Section \ref{homogeneoustreatment}, we solve Einstein equations with matter in the bulk, and the consequences that it has on the brane. In particular, we solve for the density profile of a 5D spherically collapsing atmosphere  and compute the change on the Hubble parameter as it falls into our 3-brane.  In Section \ref{curvature.perturbation}, we study cosmological perturbations in the bulk and their projection onto the brane, making special emphasis on the curvature perturbation and its power spectrum. Section \ref{data} compares our predictions against Planck data and contrasts it with the power-law power spectrum assumed in the $\Lambda$CDM model. We conclude our work with discussion of the limitations and prospects of our model in Section \ref{conclusions}.


\section{A 5D Holographic Big Bang} \label{strategy}

\subsection{Brane Nucleation} \label{nucleation}

As described in the introduction, we are working in the context of the 5D Holographic  
Big Bang model  \cite{Holobigbang2013} where our universe is modelled
as a hypersurface (the brane) in a 5-dimensional Schwarzschild space time 
according to the embedding $r=\frac{a(\tau)}{\sqrt{{\cal{K}}}}$. This 
construction is a solution of the DGP action  \eqref{DGPaction} once we impose a
$\mathbb{Z}_2$ boundary condition on the brane. 
As a consequence, via the embedding constraint the brane becomes an outward travelling boundary of the higher dimensional spacetime, an assumption that is necessary 
if we want our universe (represented by the brane) to be expanding.  

From the perspective of an observer in the bulk, this setup is
reminiscent of a construction proposed by Witten called the {\it{`bubble of nothing'}} \cite{Witten1982481}, 
in which an interior region of space is missing, with space ending smoothly at the surface of this 
bubble (the brane).  
One possible scenario in the 5D Holographic Big Bang model is that the brane was formed by the
quantum tunnelling of a collapsing star in 5 dimensions, with all of the degrees of freedom 
of the inner part of the collapsing matter becoming degrees of freedom of the brane.  This 
is analogous to  the {\it fuzzball} paradigm, a model proposed to solve the  information-loss 
paradox \cite{Lunin:2002qf}, which consists
of the explicit construction of black hole microstates with no ``dataless horizon 
region''. The infalling matter can tunnel to a fuzzball state with amplitude
\cite{Mathur:2008kg}
\beq
{\cal{A}}\sim e^{-\frac{1}{G_5}\int R}\sim e^{-\alpha G_5 M^3}\,,
\eeq
where $\alpha={\cal{O}}(1)$ and we have used the length scale $r\sim G_5 M^2$ to estimate the Euclidean Einstein action for tunneling between two configurations that have the 
length and mass scales set to those of the black hole.  Although this amplitude is very small, 
 the number of fuzzball configurations that a black hole can tunnel to 
depends on the Bekenstein-Hawking entropy as
\beq
{\cal{N}} \sim e^{S_{BH}}\sim e^{G_5M^3} 
\eeq
yielding a significant probability to form a fuzzball. In fact, the two exponentials exactly cancel each other \cite{Mathur:2008kg}.  We anticipate a similar principle operating here, in which collapsing  matter at sufficiently high density -- just prior to formation of a black hole horiozn-- 
necessarily tunnels to a brane so as to avoid the ensuing quantum paradoxes that follow upon introducing an event horizon. 

Since it is well established that BBN happened in the formation of our universe, and that in the 5D Holographic Big Bang (HBB) model \cite{Holobigbang2013}
 the pressure singularity generically forms before BBN, we consider a 
brane that  must  form before this. This means that the temperature of nucleation must be at most the temperature of BBN --  $T_{\rm BBN} \sim 0.4\,\text{MeV}$ \cite{Pospelov:2010hj}:
\beq
T_{\rm nuc} \geq T_{\rm BBN} \sim 0.4 ~{\rm MeV}.
\label{BBNconstrain}
\eeq
Finally, let us mention that the DGP model possesses a scale $r_c=\frac{G_5}{G_4}$,
above which  5-dimensional gravity dominates over  4-dimensional gravity.
Constraints on the normal branch of the DGP model \cite{Azizi:2011ys} give \footnote{Planck 2015 constraints on the dark energy equation of state roughly imply $|1+w| < 0.11$ at 95\% level, at the pivot redshift of $z \simeq 0.23$ (Fig. 5 in \cite{Ade:2015rim}), which provides a similar bound on $r_c$, using the DGP Friedmann equation with a cosmological constant.}
\beq
r_c\gtrsim3H_0^{-1}\,\,\,\,\,\,\,\,\, \to\,\,\,\,\,\,\,\, M_5<(H_0 M_4^2/12)^{1/3}\,\,\,\,\,\,\,\,
\to\,\,\,\,\,\,\,\,M_5< 9\,\text{MeV}\,.
\label{DGPconstrain}
\eeq
where $M_4=\frac{1}{(8\pi G_4)^{1\!/2}}$ is the reduced 4D Planck mass. 

%
%


\subsection{The Atmosphere: Setup and Scales}

\begin{figure}
\centering
  \includegraphics[scale=0.4]{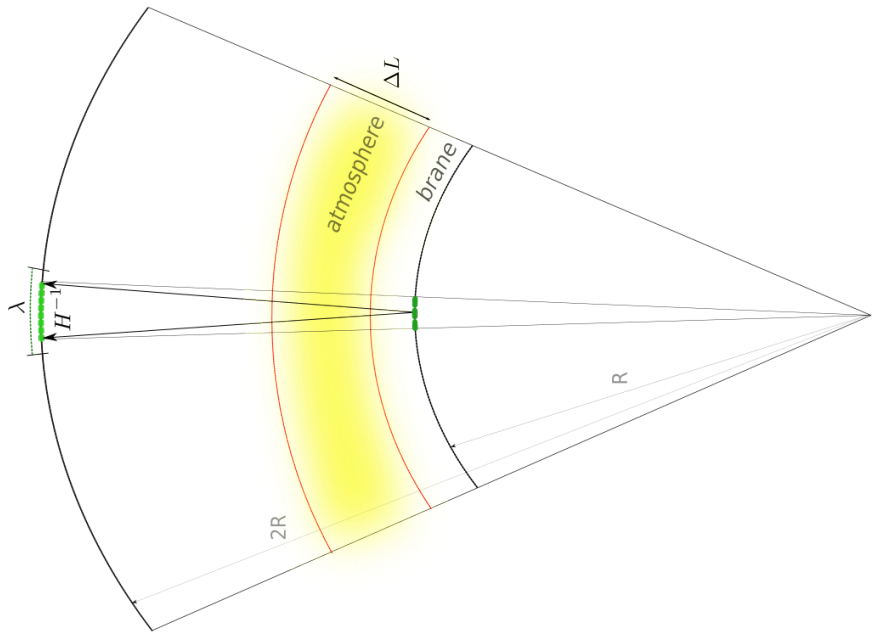}
  \caption{Cartoon of the different scales treated in this problem, in the black hole rest frame. The inner black circular arc represents the brane with radius $R$
  and the outer black circular arc represents the brane with radius $2R$. $H^{-1}$ is estimated by tracing light rays on the brane after
  it has doubled it size; it is small if the brane is traveling near the speed of light. The atmosphere is shown in yellow sitting in 
  between the two red arcs with length $\Delta L$.}
  \label{fig.scales}
  \end{figure}
  
In this scenario, we are interested in the effects of a thin atmosphere located just outside the 
brane. Although the brane forms
a  $\mathbb{Z}_2$ boundary, excluding the event horizon in Eq. (\ref{bulk_sch}), we shall
refer to the metric in Eq. (\ref{bulk_sch}) as the black hole metric. 

In order to organize the different assumptions, we will review the implied hierarchy of
the scales present in this problem. If we assume that the Hubble patch of our universe, at/near brane nucleation, 
is small enough to be insensitive to the curvature of the black hole spacetime, we can 
assume that the atmosphere
is just a perturbation around a Minkowski background. This limit implies $H^{-1}\ll R$ where R is the radius of the
black hole, or brane upon nucleation. This assumption is also observationally motivated, since today we measure $(HR)^{-2} \ll |\Omega_k|\ll1$ and thus our observable Hubble patch is approximately flat.
In our model, if the brane is moving very close to the speed of light, $H^{-1} \ll R$. 
This will  allow us to define metric perturbations in Section \ref{curvature.perturbation}. We will then be interested in finding the behaviour of the power spectrum 
of curvature perturbations for  modes  of
 wavelength $\lambda$, that are of super-horizon size before BBN, but are now observable in the CMB sky. This restricts  $R\gg \lambda \gg H^{-1}$. 

Finally, we would like to understand the behaviour of different physical quantities of the brane, such as the behaviour of the Hubble 
parameter before and after the encounter with the atmosphere.  Assuming it  can be considered to be a  thin atmosphere in the 5
dimensional space time, the width $\Delta L$ of the atmosphere 
needs to be smaller than $R$ for the brane to cross the atmosphere completely in less than a Hubble time.
We then get the following hierarchy of   different scales 
\bea
 H^{-1} \ll \lambda  \ll R, ~{\rm and}~ \Delta L \ll R {\rm ~(black~hole~frame)}, \\
 \lambda \lesssim \Delta L {\rm~(atmosphere~frame)}.
 \label{scales_constrain}
\eea
As we shall in Section \ref{curvature.perturbation}, the latter inequality is the key ingredient that leads to a near scale-invariant spectrum of primordial curvature perturbations for large scales.
This hierarchy of scales is illustrated in Fig. \ref{fig.scales}.


 \section{Homogeneous brane meets thin atmosphere}\label{homogeneoustreatment}

\subsection{Einstein Equations}

\begin{figure*}\label{penrosediagram}
\minipage{0.4\textwidth}
  \includegraphics[width=\linewidth]{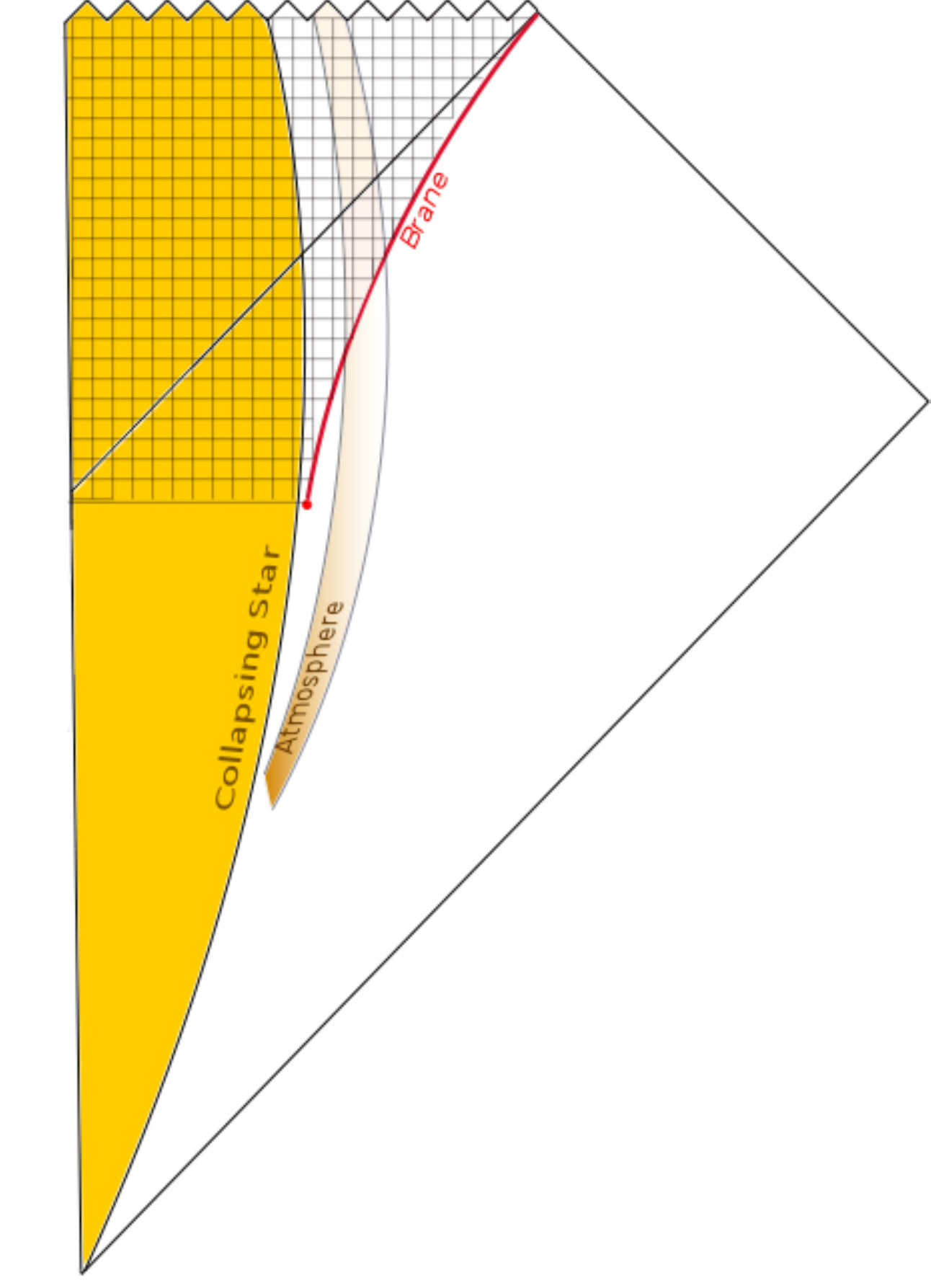}
  \endminipage\hfill
\minipage{0.5\textwidth}
  \includegraphics[width=\linewidth]{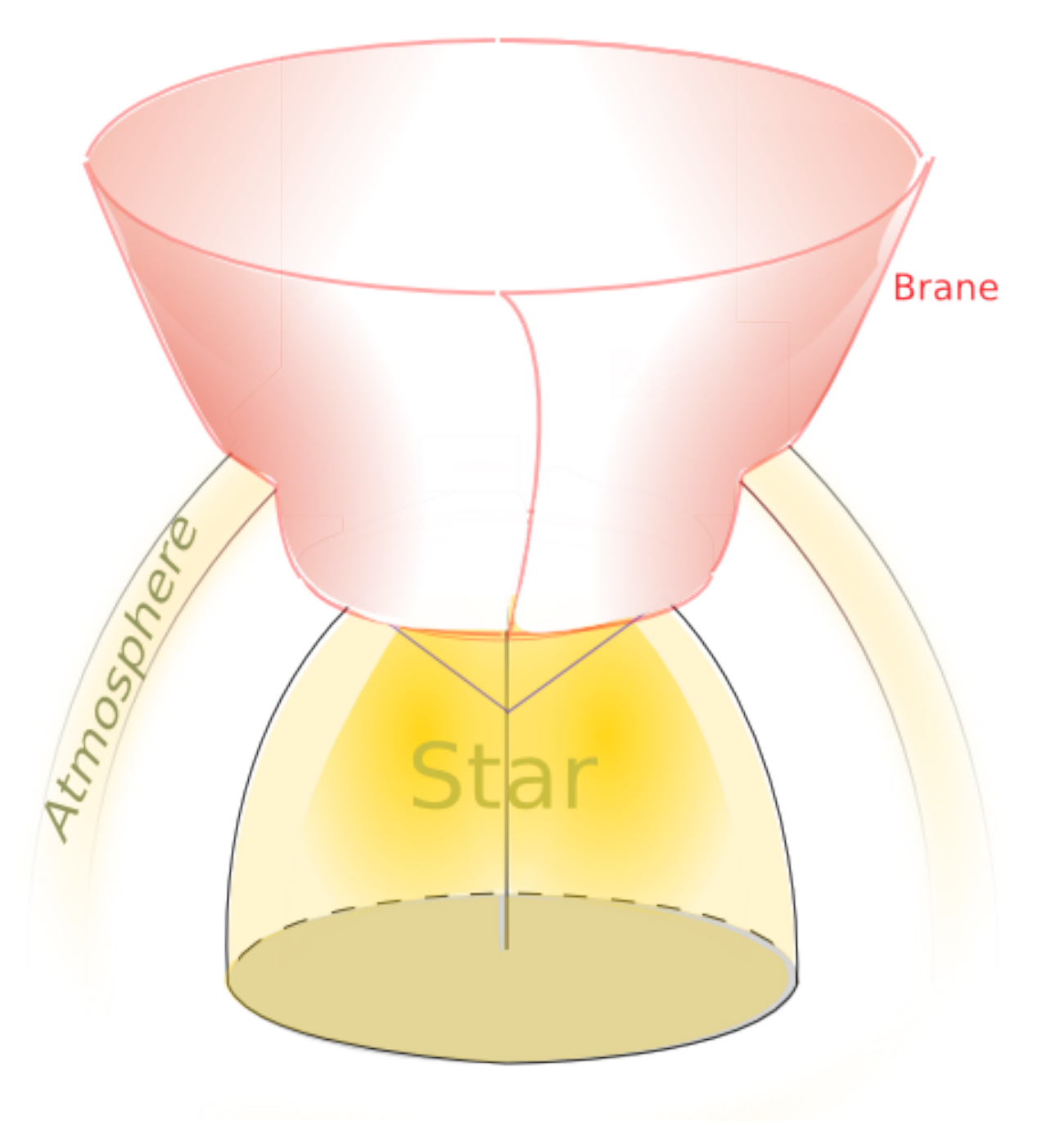}
  \endminipage
  \caption{Penrose diagram (left), and cartoon of the 5D star collapse (right), followed by the nucleation of a 3-brane (our universe). The star (in yellow) that is collapsing
(nearly) forms a black hole, but the 3-brane (red) will  nucleate just prior to the formation of the event horizon, and traverses a thin atmosphere of infalling matter or atmosphere
(cross-section shown in the cartoon at right).}
\end{figure*}

We want to study the influence of an  atmosphere that is falling into
the black hole as shown in in Fig. \ref{penrosediagram}. 
For this, we assume that the brane is moving supersonically (in fact, almost with the speed of light) into the atmosphere, and thus bulk metric perturbations do not react to the brane's presence until it runs into them.

The Einstein equations on the brane 
that follow from the action \eqref{DGPaction1}
are:
\beq
G_{\mu\nu}=8\pi G_4\left(T_{\mu\nu}+\tilde{T}_{\mu\nu}\right) 
\label{eint_tensor_brane}
\eeq
where the two different components of the energy-momentum tensor are $T_{\mu\nu}$,
the matter living on the brane, and $\ti{T}_{\mu\nu}$ the {\it{holographic fluid}} that is induced 
on the brane via the junction conditions described below.
 Due to the symmetry of the (unperturbed) FRW spacetime, $T_{\mu\nu}$ has the form of a perfect fluid
 \beq
 T_{\mu\nu}=\argu{P+\rho}u_\mu u_\nu +P\gamma_{\mu \nu}\,,
 \eeq
where $u^{\mu}$ is the 4-velocity of the fluid normalized such that $u^{\mu}u_{\mu}=-1$.

The holographic fluid is the Brown-York stress tensor induced on the brane once 
Einstein equations are imposed on the bulk  
\beq
\tilde{T}_{\mu\nu}=\frac{1}{8\pi G_5}\argu{K\gamma_{\mu\nu}-K_{\mu\nu}}, 
\label{holoextrin}
\eeq
where $K_{\mu\nu}\equiv \nabla_{\alpha}n_{\beta}e^{\alpha}_{\mu}e^{\beta}_{\nu}$ is the 
extrinsic curvature of the brane whose unit  normal is $n^{\alpha}$. Here  $e^{\alpha}_\nu \equiv \frac{\partial 
\hat{x}^{\alpha}}{\partial x^{\nu}}$, where we have associated
the set of coordinates  $\{\hat{x}^{\alpha}\}$ and $\{x^{\nu}\}$ with the bulk and
the brane respectively.
In addition to the Einstein equations \eqref{eint_tensor_brane}, the continuity
equations for the total matter living on the brane arising from the Bianchi identities
are:
\beq
\nabla^{\mu}\argu{T_{\mu\nu}+\tilde{T}_{\mu\nu}}=0\,.
\label{4conser}
\eeq
The Gauss-Codazzi equations constrain the geometric quantities of the brane with 
 the matter present in the bulk 
\bea
\nabla^{\mu}\argu{Kg_{\mu\nu}-K_{\mu\nu}}&=&8\pi G_5 T^5_{\alpha \beta}e^{\alpha}_{\nu}n^{\beta}\,, \label{gc1} \\
R^4+K^{\mu \nu}K_{\mu \nu}-K^2&=&-16 \pi G_5 T^5_{\alpha \beta}n^{\alpha}n^{\beta}\,, \label{gc2}
\eea
where $R^4=-8\pi G_4 \argu{T+\tilde{T}}$   is the Ricci scalar of the brane. 
 $T^5_{\alpha \beta}$ 
is the energy momentum tensor of the bulk which satisfies the Einstein's equations 
on the bulk $G_{\alpha \beta}=8\pi G_5 T^5_{\alpha \beta}$. Note that the first of the Gauss-Codazzi equations (\ref{gc1}) reduces to the conservation of the holographic fluid $\tilde{T}_{\mu\nu}$ in the case that
$T^5_{\alpha \beta}=0$. If the bulk matter flows into the brane, the holographic fluid 
is not conserved and the effect of the continuity equation \eqref{4conser}
is to change the matter on the brane through the holographic fluid in order
for the sum of both to be conserved. 

In the same way, as a result of the symmetries of FRW spacetime, the holographic 
fluid must have the form of a perfect fluid. Moreover, the 4-velocity of this 
fluid must coincide with the 4-velocity 
of the normal matter on the brane:
\beq
\tilde{T}_{\mu\nu}=\argu{\tilde{P}+\ti{\rho}}u_{\mu}u_{\nu}+\ti{P}\gamma_{\mu\nu}\,. 
\label{holoperfect}
\eeq
Combining Eqs. \eqref{holoextrin} and \eqref{holoperfect} we get:
\beq
K_{\mu\nu}=-8\pi G_5\bigg[\argu{\ti{P}+\ti{\rho}}u_{\mu}u_{\nu}+\frac{1}{3}\ti{\rho}_{\mu\nu}\bigg]\,.
\eeq

\subsection{Shift in the Hubble} \label{sec:shiftH}

The rate of expansion described by the Hubble parameter will change as the brane 
goes through the atmosphere and we can find a general expression for $H$ by studying the Einstein equations and junction conditions for the DGP brane in the general case. This general case 
treats the   bulk as a 5-dimensional Schwarzschild black hole \eqref{bulk_sch} and the brane as a hypersurface parametrized by $r=\frac{a(\tau)}{\sqrt{{\cal{K}}}}$, as detailed
in the Introduction. 

From Equations \eqref{eint_tensor_brane} and (\ref{4conser}-\ref{gc2}) we obtain
\bea
&&H^2+\frac{\cal{K}}{a^2}=\frac{8\pi G_4}{3}\argu{\ti{\rho}+\rho} \label{eq.1} \\
&&\dot{\rho}+\dot{\ti{\rho}}+3H\argu{\rho +\ti{\rho}+P+\ti{P}}=0 \label{eq.2} \\
&&\dot{\ti{\rho}}+3H\argu{\ti{\rho}+\ti{P}}=T^5_{\alpha \beta}e^{\alpha}_{\tau}n^{\beta} \label{eq.3}\\
&&0=T^5_{\alpha \beta}e^{\alpha}_{i}n^{\beta} \label{eq.4}\\
&&\argu{\rho +\ti{\rho}}-3\argu{P+\ti{P}}+\frac{8\pi G_5^2}{G_4}\argu{\frac{2}{3}\ti{\rho}^{\,2}+2\ti{P}\ti{\rho}}=-2\frac{G_5}{G_4}T^5_{\alpha \beta}n^{\alpha}n^{\beta} \label{eq.5}
\eea
The quantity $T^5_{\alpha\beta}$ is the stress-energy of  the atmosphere outside the black hole.
This atmosphere will have two effects on the brane. It will induce metric and matter 
perturbations in our universe, which we shall use to compute the curvature perturbation in Sec. \ref{curvature.perturbation} below.
However it will also make the Hubble parameter change its value as the brane
crosses the atmosphere:  the brane will expand more slowly  
due to an extra source of infalling matter. Combining Eqs. \eqref{eq.1} and \eqref{eq.2}  and ignoring the curvature term we get 
\beq
\Delta H = -4\pi G_4 \int (P_T+\rho_T)\, d\tau,  \,
\label{deltaH1}
\eeq
where the integral is performed in the proper time of the brane and $P_T = P+\ti{P}$, $\rho_T = \rho +\ti{\rho}$.

Let's first look at the behavior of the holographic fluid $\ti P$ and $\ti \rho$. 
From Equation 
\eqref{eq.5} we have 
\beq
\ti P+ \ti \rho =\frac{1}{3\big(\frac{\ti \rho}{\ti \rho_s}-1\big)}\bigg[-2\frac{G_5}{G_4}
T^5_{nn}-4\ti \rho +2 \frac{\ti \rho^2}{\ti \rho_s}+ T\bigg]\,,
\eeq
where $T=3P-\rho$, $T^5_{nn}=T^5_{\alpha\beta} n^{\alpha}n^{\beta}$ and $\ti \rho_s=\frac{3G_4}{16\pi G_5^2}$. In order to avoid the pressure singularity we require
$\ti \rho \gg \ti \rho_s$ and in this limit the last equation becomes
\beq
\ti P+ \ti \rho=\frac{1}{3}\bigg[-2\frac{G_5}{G_4}
T^5_{nn}\frac{\ti \rho_s}{\ti \rho} +2\ti \rho +T \frac{\ti \rho_s}{\ti \rho}\bigg]\,.
\label{holofluideq}
\eeq

Now let's analyze the behavior for the the matter on the brane $P$ and $\rho$. Combining Eqs. \eqref{eq.2} and \eqref{eq.3}
and assuming an equation of state $P=w\rho$ the fluid on the brane satisfies
\begin{equation}
 \dot{{\rho}}+3H\argu{{\rho}+{P}}=-T^5_{\alpha \beta}e^{\alpha}_{\tau}n^{\beta} \Rightarrow
\frac{d}{d\tau}(\rho a^{3(w+1)})=-T^5_{\alpha \beta}e^{\alpha}_{\tau}n^{\beta} a^{3(w+1)}\,.
 \end{equation}
If we now assume that the atmosphere is thin enough so that we can approximate the matter distribution as a delta function (i.e. $H\Delta\tau \ll 1$, during the impact time $\Delta\tau$), we see that 
the last equation will give a jump in the density (and hence in the pressure) proportional to a step function. In fact, if we consider the system of equations (\ref{eq.1}-\ref{eq.5}), with $P=w\rho$, and a delta function $T^5_{\alpha\beta}$, the only consistent solution would have a delta function in $\tilde{P}$, with step function jumps in other variables. 
As such, the biggest contribution in Eq. \eqref{deltaH1} is given by the first term on the right hand side of Eq.  (\ref{holofluideq}):
\beq\label{dH-eq}
\Delta H=\frac{G_4}{2G_5}\int \frac{T^5_{\alpha\beta}n^{\alpha}n^{\beta}}{\ti{\rho}} d\tau \left[1+ {\cal O}(H\Delta\tau)\right] \,.
\eeq
\\
We shall see in Sec. \ref{curvature.perturbation} below that the amplitude of curvature perturbations depends on $\Delta\ln H=\frac{\Delta H}{H}$. To compute this, we note that from Eq. \eqref{eq.1} we 
can write $H^2 \approx \frac{8\pi G_4}{3}(\tilde{\rho}+\rho)\approx\frac{8\pi G_4}{3}\rho$ in the regime where
$\rho \gg \ti \rho$. 
 Then 
\beq
\Delta\, \text{ln}H \approx \frac{G_4}{2G_5}\sqrt{\frac{3}{8 \pi G_4}}\frac{1}{\sqrt{\rho}\ti \rho}
\int T^5_{\alpha\beta}n^{\alpha}n^{\beta}\, d\tau\,,
\label{logH1}
\eeq
\\
We can now use the solution in vacuum for $\ti \rho$ found in \cite{Holobigbang2013}
\beq
\ti \rho =\ti \rho_s\bigg(1+\sqrt{1-\frac{2(\rho_{BH}-\rho)}{\ti \rho_s}}\bigg)\,,
\eeq
where $\rho_{BH}=\frac{3\Omega_k^2H_0^4r_h^2}{8 \pi G_4 a^4}$. In the approximation where $\rho\gg \rho_{BH}$
and $\rho \gg \ti \rho_s$ we find
\beq
\ti \rho \approx \rho \sqrt{\frac{2\ti \rho_s}{\rho}}\,
\eeq
and then Eq.~\eqref{logH1} reads
\beq
\Delta\, \text{ln}H \approx \frac{1}{2\rho}\int T^5_{\alpha\beta}n^{\alpha}n^{\beta} d\tau\,,
\label{logH}
\eeq
  implying that the relative jump in  the Hubble parameter due to a thin atmosphere is the ratio of the work done by the pressure of the atmosphere to the energy of the brane.

\subsection{Profile of the atmosphere}\label{sec:profile}

So far we have considered a general energy momentum tensor on the bulk responsible of dynamic features on the brane. Let's now consider that the bulk is filled with a  relativistic spherically symmetric, collapsing 5D radiation atmosphere ($P_5=\frac{1}{4} \rho_5$) that represents the atmosphere whose energy momentum tensor is
\beq
T_{\alpha\beta}^5(w)=\rho_5(w)\bigg[\bigg(1+\frac{1}{4}\bigg)\delta^0_{\alpha}\delta^0_{\beta}+\frac{1}{4}\eta_{\alpha\beta}\bigg]\,.
\label{energymomentum5}
\eeq
In order to study the effect of this atmosphere we need to introduce scalar homogeneous perturbations in the bulk. A generalization of 4D perturbation theory allows us to write the perturbed metric of the bulk in the Newtonian gauge in 5D  as
\beq
ds^2_{\text{bulk}}=-[1+2\Phi_5(x^{\alpha})]dt^2+[1-2\Psi_5(x^{\alpha})][dx^2+dy^2+dz^2+dw^2]\,.
\label{bulkmetric}
\eeq
where $\Phi_5$ and $\Psi_5$ represent the scalar perturbations of the bulk and $x^{\alpha}$
are bulk coordinates. Our universe is represented as a hypersurface in the 5D bulk, whose trajectory is 
given by the constraint $w=f(x^\mu)$. In this setup, the brane will inherit three bulk 
coordinates $\{t,x,y,z\}$
and  will respond to perturbations that are just functions of the bulk time via 
the relation $w=f(x^\mu)$. Consider that the metric perturbations and the brane position are homogeneous:
\beq
\Phi_5=\epsilon\Phi_5^0(w)\,, \quad \Psi_5=\epsilon\Psi_5^0(w)\,, \quad f=f_0(t)\,,
\label{homopert}
\eeq
where $\epsilon\ll1$ is a parameter that controls the homogeneous metric perturbations.
Assuming hydrostatic equilibrium in the (infalling) rest frame of the atmosphere, the Einstein Equations in the bulk for the metric \eqref{bulkmetric} leads to the relativistic Poisson and Euler equations:
\bea
\nabla^2 \Phi_5^0&=&\frac{8\pi G_5}{3}\rho_5\,,\\
\nabla \Phi_5^0&=&-\frac{1}{4}\frac{\nabla \rho_5}{\rho_5}\,.
\eea
These equations can be  solved exactly in Minkowski background:
\beq
\rho_5(w)=\bar{\rho}_{5}\bigg\{1-\tanh^2\bigg[\bar{\gamma}\bigg(\frac{w}{\bar{w}}-1\bigg)\bigg]\bigg\}\,,
\label{atmosphere_density}
\eeq
where $\bar{\rho}_5\equiv \frac{3}{16\pi G_5}\frac{\bar{\gamma}^2}{\bar{w}^2}$, while $\bar{\gamma}$ and $\bar{w}$ are constants of integration.

The relationship between the energy density and the temperature of the atmosphere can be computed by integrating the Bose-Einstein distribution in 4+1 dimensions
\beq
\rho_{5}(w)=\int\frac{d^4k }{(2\pi)^4}\frac{\omega}{\exp\left[\omega/T_5(w)\right]-1}=\frac{3\zeta_R(5)}{\pi^2}T_5(w)^5\,,
\label{atmos_temperature}
\eeq
where $\omega^2=k_{\alpha}k^{\alpha}$. The above expression allows us to write
\beq
T_5(w)=\bigg(\rho_5(w) \frac{\pi^2}{3\zeta_R(5)}\bigg)^{\frac{1}{5}}=1.26\,\rho_5(w)^{1/5}\,.
\label{atmos_temperature1}
\eeq
Note that the characteristic thickness of the atmosphere is given by
\beq
\Delta L = \frac{\bar{w}}{\sqrt{2} \bar{\gamma}}\,.
\label{atmos_amplitude}
\eeq

We are now in position to compute Eq. \eqref{logH} and we will do so in the reference frame of the atmosphere. In this  frame $\rho_5$ does not depend on time, but the normal to the brane $n^{\alpha}$ 
will depend on the relative velocity of the brane and the atmosphere. First consider the fluid velocity $u^\alpha=(1,{\bf{v}})/\sqrt{1-v^2}$, where {\bf{v}} is the relative 3 velocity between the brane and the atmosphere. If we now  require $n_\alpha n^\alpha=1$ and $n^\alpha u_\alpha=0$ we have  $n^\alpha=(v,{\bf v}/v)/\sqrt{1-v^2}$. 
With this we can write $T^5_{\alpha\beta}n^{\alpha}n^{\beta}=\rho_5(w)(1+4v^2)/4(1-v^2)$ and the RHS of Eq. \eqref{logH} reads
\begin{eqnarray}
\Delta\, \text{ln}H=\frac{1}{2\rho}\int T^5_{\alpha\beta}n^{\alpha}n^{\beta} d\tau&=&\frac{1}{2\rho}\int_0^\infty T^5_{\alpha\beta}n^{\alpha}n^{\beta} \sqrt{1-v^2} dt\,, \nonumber \\
&=&\frac{1}{2\rho}\int_0^\infty T^5_{\alpha\beta}n^{\alpha}n^{\beta} \sqrt{1-v^2}\frac{dw}{v}\,, \nonumber \\
&=&\frac{(1+4v^2)}{v\sqrt{1-v^2}}\frac{ \bar{\rho}_5}{8 \rho}\frac{\bar{w}}{\bar{\gamma}}\bigg[1+\tanh(\bar{\gamma})\bigg]\,. \label{constrainonlogH}
 \end{eqnarray}

%
%
%
%


\section{Cosmological perturbations} \label{curvature.perturbation}

As discussed in the last section,
if the velocity of the brane is near the speed of light we can assume that the
Hubble patch of the universe will be smaller than the curvature radius of the black hole spacetime. In this regime it is safe
to approximate the bulk as Minkowski spacetime and analyze the perturbations around it. In the last section we have briefly introduced the homogeneous scalar perturbations and in Appendix \ref{app:perturbations} we present the anisotropic perturbations. The curvature perturbation can be written as function of the scalar gauge invariant quantities \eqref{4metper1}
\beq
\zeta =  \Psi_4 -\frac{H}{\dot{H}}\bigg(H \Phi_4 + \dot{\Psi}_4\bigg)\,.
\label{curv.pert}
\eeq
Note that in our framework the Hubble constant is of first order in the perturbation (see Eq.\eqref{perturbed_hubble}) as the brane crosses the atmosphere, and thus the term $H\Phi_4$ can be neglected. With this we have 
 \beq
 \zeta \approx  \Psi_4 - \frac{\Delta \Psi_4}{\Delta \text{ ln}(H)}\,,
 \label{cur_pert-zeta}
\eeq 
where $\Delta \Psi_4=\Psi_{4_f}-\Psi_{4_i}$. Here $\Psi_{4_{i(f)}}$ stands for the metric perturbation
in 4D right before (after) the brane crossed the atmosphere, and we assume that $\zeta$ evolves continuously.
We are interested in the value of the  curvature perturbation after the brane has passed 
through the atmosphere where the metric perturbation $\Psi_{5_f}=0$, which makes $\Psi_{4_f}=0$.
In this regime the curvature perturbation becomes
\begin{equation}
 \zeta = \zeta_f \approx  \frac{\Psi_{4i}}{\Delta \text{ ln}(H)}\, .
 \label{cur_pert}
\end{equation}

 We are now ready to analyze the behaviour of the curvature perturbation power spectrum  
\beq
P_{\zeta}(k)=\int d^3{\bf{x}} \,\,e^{i\bf{k\cdot x}}\langle\zeta(x)\zeta(0)\rangle\,.
\label{powerspectrum}
\eeq
If the  atmosphere is not in
 thermal equilibrium, we can relate the 2-point correlation function of the thermal fluctuations in 5D energy density to the temperature profile of the atmosphere:
\beq
\langle\rho_5(y_1)\rho_5(y_2)\rangle=\alpha \,(T_5(y_1))^6\delta^4(y_1-y_2)\,.
\eeq
 To proceed, we notice that in \cite{Holobigbang2013} it was found that the 5D energy density correlation function due to a thermal gas is 
 \beq
 \submin{\rho_5(y_1)\rho_5(y_2)}\simeq\frac{5}{8}\bigg[\int \frac{d^4k }{(2\pi)^4}
 \bigg[\frac{1}{\text{exp}(\omega/T_5)-1}+\frac{1}{2}\bigg]\,\omega\, \text{exp}[ik_a(y_1^a-y_2^a)]\bigg]^2\,.
 \label{eq:cor}
 \eeq
 This expression can be approximated 
 as a delta function $\submin{\rho_5(y_1)\rho_5(y_2)}\simeq
  \alpha (T_5(y_1))^6 \delta^4(y_1-y_2)$, where $\alpha=\frac{5}{8}\bigg[\frac{1}{\pi^2 63}(945\zeta_R(5)-\pi^6)\bigg]$, while we have dropped the power-law UV divergence and
  $\zeta_R$ is the Riemann zeta function.  Note, that \eqref{eq:cor} can be approximated by a 4 dimensional delta function  on length scales larger than the thermal wavelength $T^{-1}_5$.
  
With the use of the Poisson equation in 5D 
\beq
\nabla^2\Psi_5(y)=\frac{8\pi G_5}{3}\rho_5(y)\,,
\label{poisson5}
\eeq
 we can find the power spectrum for the curvature perturbation to be (see Appendix \ref{app:power} for details)
 \begin{eqnarray}
{\cal{P}}(k)=\frac{k^3}{2\pi^2}P_{\zeta}(k)&=&\beta\, k \int_0^{\infty} dw\, e^{-2|w|k}\,(T_5(w))^6\\
&=&\Delta_0^2\, k \int_0^{\infty} dw\, e^{-2|w|k}\,\left\{1-\tanh^2[\bar{\gamma}(w/\bar{w}-1)]\right\}^{6/5}
\label{normalizedps}
\end{eqnarray}
  where $\Delta_0^2=\beta\,[\bar{T}_5]^6$, $\beta=\frac{\alpha}{2}\bigg(\!\frac{G_5}{6\Delta \text{ln}H \,\pi^3}\!\bigg)^2$ and we have  used Eqs.\eqref{atmos_temperature} and \eqref{atmos_temperature1} of Sec.\,\ref{sec:profile} to write the temperature of the brane. The power spectrum predicted by our model is characterized by 3 free parameters  ${\bar{\gamma},\bar{w},\Delta_0^2}$ which we are going to fit to Planck data in the next section.

\section{Observational constraints on 5D holographic big bang}\label{data}
The standard model of cosmology, $\Lambda$CDM, is described by 6 parameters $(\Omega_bh^2,\Omega_ch^2,\theta,\tau,\Delta_0^2,n_s)$, the baryon density, dark matter density, angular size of the sound horizon at recombination, the optical depth to reionization, amplitude of the scalar power spectrum and its tilt respectively. This model characterizes early universe cosmology via the power spectrum of the curvature perturbations
\beq
{\cal{P}}(k)=\Delta_0^2\bigg(\frac{k}{k_*}\bigg)^{n_s-1}\,,
\eeq
where $k_*=0.05/\text{Mpc}$ is the comoving pivot scale. This form of the power spectrum, expected in slow-roll inflationary models, best fits  the CMB data with parameter values \cite{Plankresults2015}
\beq
\Delta_0^2=(2.196 \pm 0.059)\times 10^{-9}\,\,\,\,\,\,\,\,\,\,\,n_s=0.9603\pm 0.0073\,.
\eeq
We would like to compare this model with the HBB model (Eq. \ref{normalizedps}) that strictly is represented by the seven parameters $(\Omega_bh^2,\Omega_ch^2,\theta,\tau,\bar{\gamma},\bar{w},\Delta_0^2)$. In order to compare   models with the same number of parameters, will also include $\Lambda$CDM with running 
$\alpha_s=dn_s/d\,\text{ln}q$.

\begin{figure*}
\includegraphics[width=0.525\textwidth]{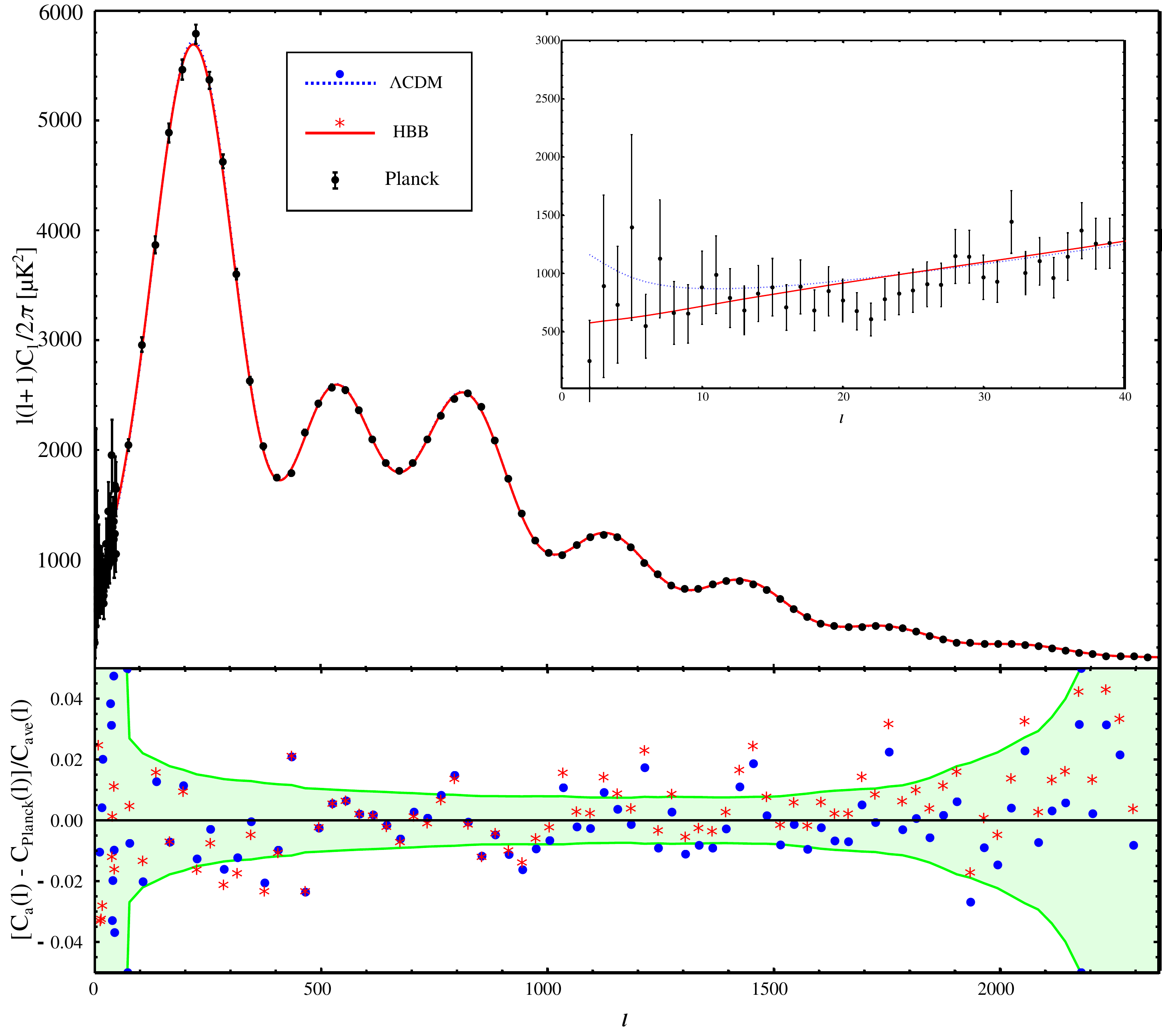}
\includegraphics[width=0.475\textwidth]{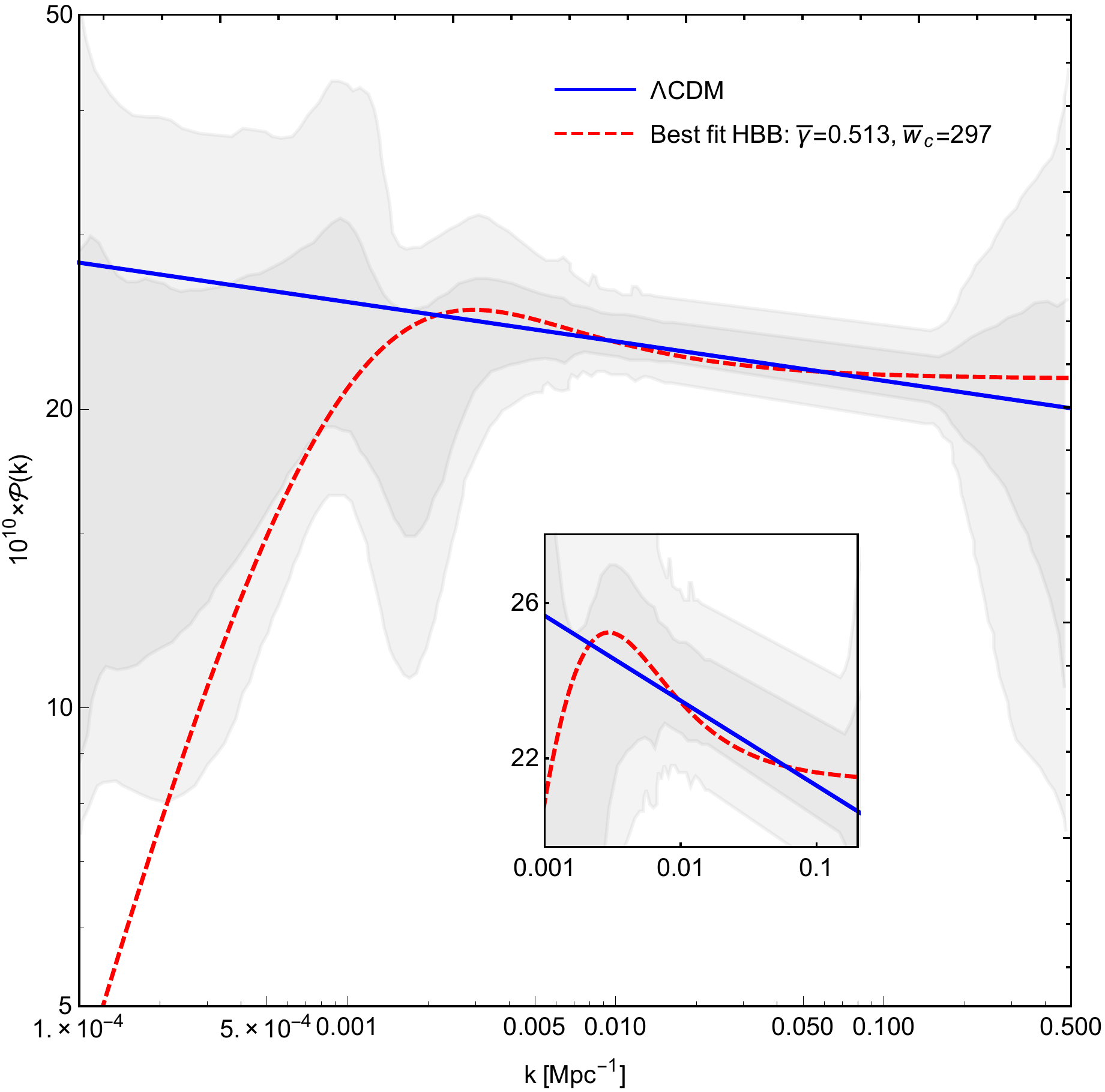}
  \caption{Left-Top: angular power spectrum of CMB temperature anisotropies, comparing Planck 2015 data (black dots) with best HBB model (solid/red) for all l. Left-Inset: angular power spectrum of CMB temperature anisotropies, comparing Planck 2015 data with $\Lambda$CDM (dotted/blue) and HBB (solid/red) for $l<40$. Left-Bottom: relative residuals and difference between $\Lambda$CDM and HBB (black solid) where the green shaded region indicates the 68\% region of Planck 2015 data. Right: Best fit of the primordial power spectrum as predicted by HBB (dashed-red) in comparison with the best fit of $\Lambda$CDM model (blue). The grey regions are the $\pm$1$\sigma$ and $\pm$2$\sigma$ constraints from a non-parametric reconstruction using Planck 2015 data \cite{Ade:2015lrj}.}
 \label{fig.plots}
\end{figure*}

We have performed the comparison by running the  CosmoMC code~\cite{1996ApJ...469..437S,1998ApJ...494..491Z,2002PhRvD..66j3511L,2000ApJ...538..473L,2012JCAP...04..027H,2013PhRvD..87j3529L,campnotes}  with Planck 2015 data, Barionic Acoustic Oscillations (BAO) \cite{2011MNRAS.416.3017B,2011MNRAS.418.1707B,2012MNRAS.427.3435A,2012MNRAS.423.3430B,2012MNRAS.427.2132P,2014MNRAS.441...24A,2014MNRAS.439.3504S,2015MNRAS.449..835R} as well as lensing data \cite{Ade:2015zua,Aghanim:2015xee,Ade:2015fva,2012ApJ...755...70R,2014JCAP...04..014D}. Finally, to determine the best fit parameters and the likelihood, we run the minimizer expressing our results Table \ref{tab:bf}. Comparing best-fit $\chi^2$ of HBB and $\Lambda$CDM (with running), we see that HBB is disfavoured at roughly 2.7 2.8$\sigma$ (2.8$\sigma$). The Planck angular TT spectrum together with the best fit curves and residuals for HBB and $\Lambda$CDM are shown in Fig. \ref{fig.plots}.   The best fit primordial scalar power spectrum in both models are also contrasted  with a non-parametric reconstruction from Planck 2015 data  \cite{Ade:2015lrj}. We note that the difference between the two models (mostly) lies within the 68\% region and the largest disagreement of the models is at low l's or k's.

\begin{table}
\caption{\label{tab:bf}
Planck 2015 and BAO best fit parameters and 68\% ranges for
HBB and $\Lambda$CDM models. The last row
indicates the $\chi^{2}$ for each of the models. Note that $\bar{w}_c$ corresponds to the comoving value of the position of the centre of the atmosphere and its related to the physical $\bar{w}$ via $\bar{w}=\bar{w}_c \frac{2.3\times10^{-10}\text{MeV}}{T_{\text{nuc}}}$.}

\noindent \centering{}

\begin{tabular}{|c||c|c||c|c||c|c|}
\hline \hline
         & \multicolumn{2}{c||}{\bf{HBB}}  & \multicolumn{2}{c||}{\bf{$\Lambda \text{CDM}$}} & \multicolumn{2}{c|}{\bf{$\Lambda \text{CDM}$ with running}}\tabularnewline
\cline{2-7} 
                  & \bf{best fit} & \bf{$68\%$ range}       & \bf{best fit}  & \bf{$68\%$ range}       & \bf{best fit}  & \bf{$68\%$ range}      \tabularnewline
\hline 
\hline 
$\Omega_{b}h^{2}$ & $0.02212$&$0.02210\pm0.00023$ & $0.02227$ & $0.02225\pm0.00020$& $0.02231$ &$0.02229\pm0.00022$\tabularnewline
\hline 
$\Omega_{c}h^{2}$ & $0.1172$ &$0.1169\pm0.0012$   & $0.1185$  & $0.1186\pm0.0012$  & $0.1184$  &$0.1186\pm0.0012$  \tabularnewline
\hline 
$100\theta$       & $1.04113$&$1.04116\pm0.00042$ & $1.04103$ & $1.04104\pm0.00042$& $1.04108$ &$1.04105\pm0.00041$\tabularnewline
\hline  
$\tau$            & $0.081$  &$0.083\pm0.014$     & $0.067$   & $0.067\pm0.013$    & $0.069$   &$0.068\pm0.013$    \tabularnewline
\hline 
$10^9\Delta_0^2$  & $5.79$  &$0.793\pm0.021$     & $5.798$   & $5.798\pm0.019$    & $5.82$   &$5.82\pm0.020$    \tabularnewline
\hline 
$n_{s}$           &          &                    & $0.9682$  & $0.9677\pm0.0045$  & $0.9682$  &$0.9671\pm0.0045$  \tabularnewline
\hline 
$\alpha_{s}$      &          &                    &           &                    & $-0.0027$&$-0.0030\pm0.0074$ \tabularnewline
\hline 
$\bar{\gamma}$          & $0.513$  &$0.525\pm0.053$     &           &                    &           &                   \tabularnewline
\hline 
$\bar{w}_c \,[\text{Mpc}]$               & $275$    & $297_{-77}^{+39}$  &           &                    &           &                   \tabularnewline
\hline 
\hline
$\chi^{2}$        &$11327.4$ &                    & $11319.9$ &                    & $11319.6$ &                   \tabularnewline
\hline  
\end{tabular}
\end{table}

We are now going to analyze the physical implications of the best fit parameters $\{\bar{\gamma},\bar{w}_c,\Delta_0^2\}$.  In particular, we are interested in the temperature and position of the atmosphere and the change on Hubble constant of the brane when crossing the atmosphere. All of these physical quantities are related to the fitted parameters, and also to the temperature of the brane at nucleation $T_{\text{nuc}}$ and the Planck mass in the bulk $M_5$. Using $\bar{w}=\bar{w}_c \frac{2.3\times10^{-10}\text{MeV}}{T_{\text{nuc}}}$ and Eqs.~\eqref{atmosphere_density}, \eqref{atmos_temperature} and \eqref{atmos_amplitude} we can find the values for the amplitude of the energy density, the temperature and the width of the atmosphere respectively. We have summarized our results in Table \ref{table:physical}.

\begin{table}
\caption{\label{table:physical}
Physical characteristics of the HBB model using the best fit parameters presented in Table~\ref{tab:bf}. The first column shows the relevant physical parameters and their definitions in terms of the best fit parameters and related quantities.The second column shows the numerical values and scaling with $M_5$ and $T_{\text{nuc}}$.  Finally, in the last column we show the limits necessary for the thin atmosphere condition   \eqref{m5_bound}. The 5th and 6th rows show the shift in the Hubble constant when crossing the atmosphere computed using perturbations and bulk background information, respectively. The 7th row presents a function constraining the velocity of the atmosphere in the bulk that can be computed by equating the results of rows 5 and 6.  Note that $T_{\text{nuc}}\geq 0.4\,\text{MeV}$ in order for the BBN constraint to be valid. }

\noindent \centering{}

\begin{tabular}{|c|c |c| c|}
\hline \hline
&& & \bf{Thin atmosphere bound}\\
&\bf{Quantity}&\bf{Value}&$M_5 \leq 8.56\!\times\! 10^{-4}\,(\frac{\text{MeV}}{T_{\text{nuc}}})^{5/21}T_{\text{nuc}}$ \\ 
&&& \eqref{m5_bound}\\ \hline \hline
&&& \\
\bf{Position of} &$\bar{w}=\bar{w}_c \frac{2.3\times10^{-10}\,\text{MeV}}{T_{\text{nuc}}}$ & $9.872 \times 10^{27}\,\frac{1}{T_{\text{nuc}}}$&  $9.872 \times 10^{27}\,\frac{1}{T_{\text{nuc}}}$ \\
\bf{atmosphere}&&& \\
 \hline 
& && \\
\bf{Density of} &$\bar{\rho}_5=\frac{3}{16\pi G_5}\frac{\bar{\gamma}^2}{\bar{w}^2}$&$1.62\times 10^{-56}\,M_5^3\,T_{\text{nuc}}^2$&$\leq1.02 \times 10^{-65} \, (\frac{\text{MeV}}{T_{\text{nuc}}})^{5/7}\,T_{\text{nuc}}^5$\\
\bf{atmosphere}&&& \\
 \hline 
 &&& \\
\bf{Temperature} &$\bar{T}_5=1.26\, \bar{\rho}_5^{1/5}$ &$ 8.75 \times 10^{-12}\,(M_5^3\,T_{\text{nuc}}^2)^{1/5}$&  $\leq1.26 \times 10^{-13}\, (\frac{\text{MeV}}{T_{\text{nuc}}})^{1/7}\,T_{\text{nuc}}$\\
 \bf{of atmosphere}&&& \\
 \hline 
 &&& \\
\bf{Width of}&$\Delta\,L=\frac{1}{\sqrt{2}}\frac{\bar{w}}{\bar{\gamma}}$&$1.35\times 10^{28}\frac{1}{T_{\text{nuc}}}$&$1.35\times 10^{28}\frac{1}{T_{\text{nuc}}}$\\ 
\bf{atmosphere}&&& \\
\hline 
\bf{Change of}& && \\
\bf{Hubble from}&$\Delta\,\text{ln}\,H=\frac{G_5 T(0)^3}{6 \pi^3 \Delta_0}(\frac{\alpha}{2})^2$&$ 4.54\times 10^{-35}(\frac{T_{\text{nuc}}}{M_5})^{6/5} $& $\geq2.18\times 10^{-31}\,(\frac{T_{\text{nuc}}}{\text{MeV}})^{2/7}$\\ 
\bf{perturbations}&Eq.~\eqref{constrainH1}&& \\
\hline 
 \bf{Change of} &$\Delta\,\text{ln}\,H=$&& \\
\bf{Hubble from}&$f(v)\frac{\bar{\rho}_5}{8\rho}\frac{\bar{w}}{\bar{\gamma}}[1+\tanh(\bar{\gamma})]$&$ 1.62\times 10^{-29} (\frac{M_5}{T_{\text{nuc}}})^3 f(v) $&$\geq2.18\times 10^{-31}\,(\frac{T_{\text{nuc}}}{\text{MeV}})^{2/7}$\\ 
\bf{backgroud}&Eqs.~\eqref{constrainH2} and \eqref{constrainonlogH}&& \\
\hline
& && \\
\bf{Velocity}&$f(v)=\frac{(1+4v^2)}{v\sqrt{1-v^2}}$&$ 2.8\times 10^{-6}\,(\frac{T_{\text{nuc}}}{M_5})^{21/5} $& $\geq2.14\times 10^{7} \frac{T_{\text{nuc}}}{\text{MeV}}$\\ 
\bf{constraint}&& &\\
\hline

\end{tabular}
\end{table}

The model still allows freedom for the parameters $\{T_{\text{nuc}},M_5,v\}$. These can be constrained using observational bounds for DGP model and by requiring consistency of our approximations. 
In particular, we have modelled the atmosphere as being thin, which is equivalent to the requirement that the time it takes the brane to cross it is less than a Hubble time:
\beq
\Delta L\frac{\sqrt{1-v^2}}{v}\leq H^{-1}=\bigg(\frac{3}{8 \pi G_4 \rho}\bigg)^{1/2} \,\,\,\,\,\,\,\ \Rightarrow \,\,\,\,\,\,\, T_{\text{nuc}}\leq \frac{v}{\sqrt{1-v^2}}2.33 \times 10^{-7}\,\text{MeV}\,, 
\label{eq:bound_thin_shell}
\eeq 
employing the fact that the energy density on the brane at nucleation time is $\rho=\frac{\pi}{30}\,g_{*}\,T_{\text{nuc}}^{4}$,  for $g_*$ effective relativistic degrees of freedom. The above bound is consistent with the BBN constraint ($T_{\text{nuc}} \geq T_{\text{BBN}}$) for velocities near the speed of light.  On the other hand,  as discussed in Section \ref{sec:shiftH} when the brane encounters the atmosphere its Hubble parameter will change as predicted by Eq.~\eqref{constrainonlogH}. Since this quantity enters in the amplitude of the power spectrum Eq.~\eqref{normalizedps} we can write 
\beq
\Delta_0^2=\beta \,\bar{T}_5^6\,,\,\,\,\,\,\,\,\,\,\,\,\,\, \beta=\bigg(\frac{G_5}{6 \pi^3 \Delta\, \text{ln}\,H}\bigg)^2
\eeq
and thus 
\beq
\Delta\,\text{ln}\,H=4.54\times 10^{-35}\bigg(\frac{T_{\text{nuc}}}{M_5}\bigg)^{6/5} \,.
\label{constrainH1}
\eeq
We also get a constraint on $\Delta\,\text{ln}\,H$ from the bulk atmosphere using Eq.~\eqref{constrainonlogH}
\beq
\Delta\,\text{ln}\,H=1.62\times 10^{-29} \,\frac{(1+4v^2)}{v\sqrt{1-v^2}} \bigg(\frac{M_5}{T_{\text{nuc}}}\bigg)^3\,,
\label{constrainH2}
\eeq
Using Eqs.~\eqref{constrainH1}  and \eqref{constrainH2} we can constrain the velocity and the speed of sound of the brane
\beq
f(v)=\frac{(1+4v^2)}{v\sqrt{1-v^2}}=2.8\times 10^{-6}\,\bigg(\frac{T_{\text{nuc}}}{M_5}\bigg)^{21/5}\,.
\label{velocity_function}
\eeq
From the above equation we notice that in order to have a real brane velocity we need to satisfy
\beq
T_{\text{nuc}}\geq 30\, M_5\,.
\label{reality_bound}
\eeq

\begin{figure*}
\centering
\includegraphics[width=0.7\textwidth]{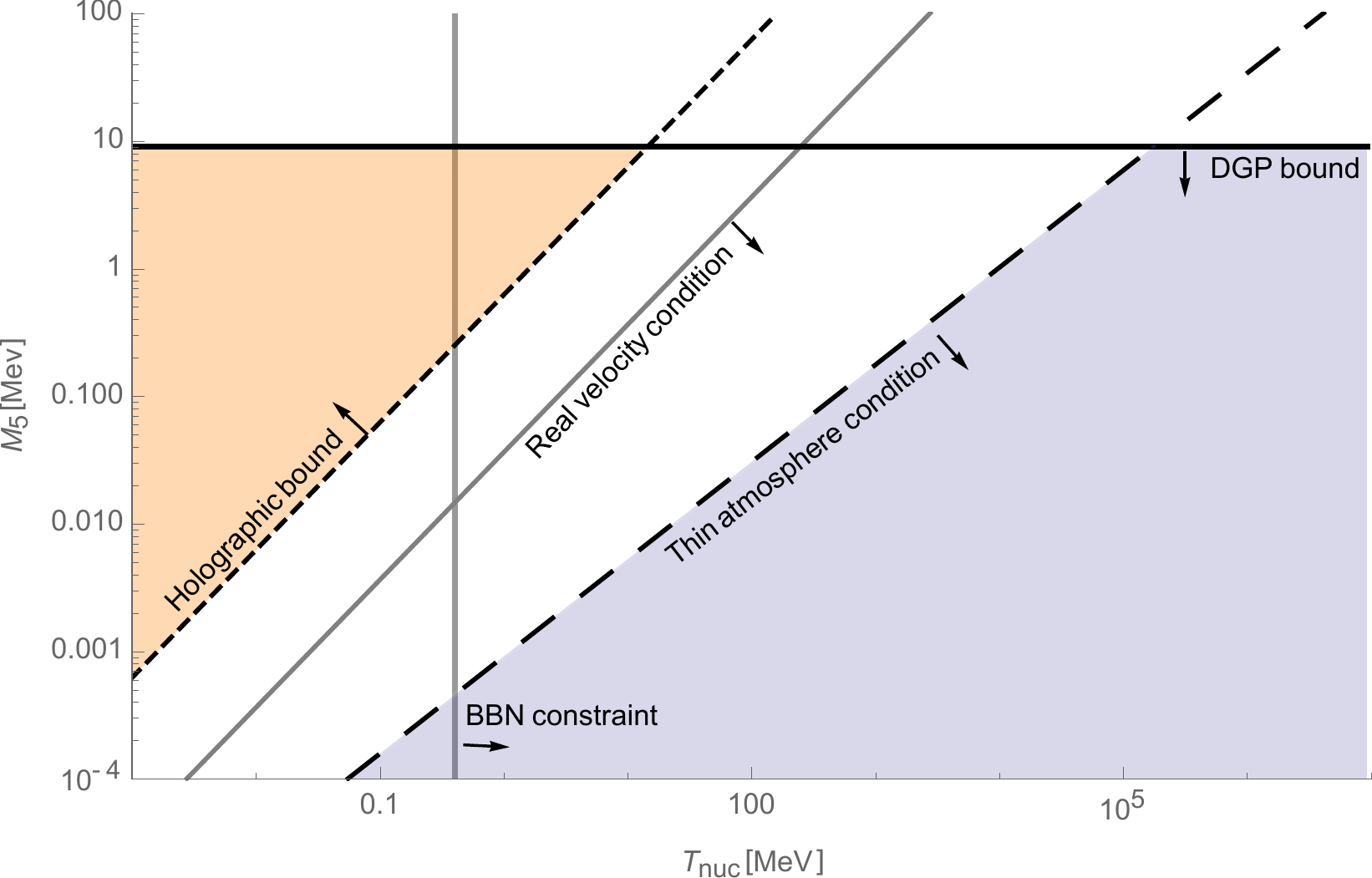}
  \caption{\label{fig:constrains}  Theoretical and empirical bounds for the Holographic Big Bang model.  The DGP bound Eq.\eqref{DGP_bound} (thick-black) and the holographic bound Eq.\eqref{entropybound} (black, thin dashed), together with the BBN bound Eq.\eqref{BBN_bound} (vertical black) constitute the theoretical bounds of the model. The top shaded area (orange) is the allowed region for these three bounds to be satisfied. The real velocity bound Eq.\eqref{reality_bound} (thick, grey) and the thin atmosphere condition \eqref{final_thin_shell} (black, thick dashed) constitute the empirical bounds that HBB must satisfy.  The bottom shaded area (blue) is the allowed region of $\{M_5,T_{\text{nuc}}\}$ parameter space satisfying the empirical bounds, and the arrows  indicate the directions in which the different bounds  apply.  It is clear  that the empirical bounds violate the holographic bound for all possible allowed pairs  $\{M_5,T_{\text{nuc}}\}$ by at least 2.5 orders of magnitude. The least severe violation of the holographic bound is for parameters at the bottom left of the plot: for $T_{\text{nuc}}$ being the minimum allowed value by BBN and $M_5$ the maximum allowed by the thin atmosphere condition (third column of Table \ref{table:physical}). }
  
\end{figure*}

%
We can obtain a constraint for $\{T_{\text{nuc}},M_5\}$ by combining expressions \eqref{velocity_function} and \eqref{eq:bound_thin_shell} and noting that in the large velocity limit $\frac{v}{\sqrt{1-v^2}}\approx\frac{1}{\sqrt{1-v^2}}\approx \frac{f(v)}{5}$ 
\beq
T_{\text{nuc}}\leq 1.31\times 10^{-13}\,\bigg(\frac{T_{\text{nuc}}}{M_5}\bigg)^{21/5} \text{MeV}\,.
\label{final_thin_shell}
\eeq
Inverting this we find 
\beq
M_5 \leq 8.56\times 10^{-4}\,\bigg(\frac{\text{MeV}}{T_{\text{nuc}}}\bigg)^{5/21}T_{\text{nuc}}\,.
\label{m5_bound}
\eeq
 This bound represents the maximum allowed value of $M_5$ in order for the thin atmosphere condition to be satisfied. In the third column of Table~\ref{table:physical} we list the values for the physical quantities allowing $M_5$ to saturate the above bound.  This constraint must be combined with the physical constraint \eqref{BBNconstrain}
\beq
T_{\rm nuc}>T_{\rm BBN} \sim 0.4 ~{\rm MeV}
\label{BBN_bound}
\eeq
as well as with the constraint \eqref{DGPconstrain} 
\beq
M_5<9\,\text{MeV}\,,
\label{DGP_bound}
\eeq
on the normal branch of DGP  in order to get the allowed region in parameter space for $\{T_{\text{nuc}},M_5\}$ -- depicted in Fig.\ref{fig:constrains} (blue shaded region). 

 It is interesting to compare the thermal entropy of our brane to the holographic bound expected from its surface area in 5D.  The entropy for the 5D black hole is 
$S_{BH}=\frac{A}{4G_5}$, while the entropy density in a universe dominated by relativistic particles
 $s(T)=\frac{4\pi^2}{90} g_*  T_{\text{nuc}}^3 \simeq 4.71 \times T_{\text{nuc}}^3$,  for $g_*= 10.75$ effective relativistic degrees of freedom, prior to electron/positron annihilation \cite{1990eaun.book.....K}.
This puts a lower bound 
\beq
M_5>0.23 ~\bigg(\frac{T_{\rm nuc.}}{0.4\,\text{MeV}}\bigg)~ \text{MeV} ~~~~~{\rm(holographic~bound)}
\label{entropybound}
\eeq
on the 5D Planck mass $M_5=\frac{1}{(32\pi G_5)^{1\!/3}}$, 
where $T_{\rm nuc.}$ is the  nucleation temperature of the brane. We show the Holographic bound allowed region in parameter space in Fig. \ref{fig:constrains} (orange shaded region).
Eqs. \eqref{entropybound}-\eqref{DGPconstrain} constrain the 5D Planck mass to be within 1.5 decades in energy:
\beq
0.23\,\text{MeV}<M_5<9\, \text{MeV},
\label{m5range}
\eeq
a range that will inevitably shrink with future observations that better constrain BBN, and late-time cosmic expansion history.  As we see in Fig.\ref{fig:constrains}, the best-fit value for $T_{\rm nuc.}$ from cosmological observations (assuming the thin atmosphere condition) does violate the holographic bound (\ref{entropybound}) by at least 2.5 orders of magnitude, 
 which would decrease the lower limit on $M_5$ in Eq. (\ref{m5range}) by the same factor\footnote{  
 Note that the exact saturation of the holographic bound predicts a brane velocity that is not real.}. 

 Is this a ``show-stopper''? While the holographic bound on entropy remains a very well-motivated conjecture, it is not clear how firm it might be as objects that get close to crossing it are already in the quantum gravity regime where the classical description of spacetime physics fails.  One may argue that since the degrees from responsible for thermal entropy of our brane are on scales much smaller than the 5D Planck length, they are not accessible by a 5D bulk observer, and thus are not limited by the 5D holographic bound.

%

%
%


\section{Summary and Discussion} \label{conclusions}

The 5D Holographic Big Bang (HBB) is a novel proposal for  a holographic origin of our universe as a 3-brane with induced gravity,  out of the collapse of 5D star that can address the traditional problems of big bang cosmology. The main goal of this study was to provide detailed and concrete predictions for this proposal, and to see whether it can serve as a possible competitor to slow-roll inflationary models to explain cosmological observations.

 We first focused our attention on a possible mechanism for the nucleation of our 3-brane in which the quantum degrees of freedom of the bulk tunnel into a fuzzball configuration reminiscent to the {\it{bubble of nothing }}model. This mechanism not just provides a possible scenario of brane nucleation but also constrains the Planck mass in the bulk. 
  
Previous work has shown that the presence of uniform thermal gas in the 5D bulk leads a scale-invariant primordial power spectrum for cosmological scalar perturbations. To formalize this result and search for mechanisms that could potentially explain deviations from scale-invariance  (observed in the CMB data), we studied cosmological perturbations induced by a thin infalling atmosphere. This atmosphere is composed of a spherically symmetric thermal relativistic gas that the brane encounters after nucleation. We showed that this atmosphere induces a change in the Hubble parameter and also scalar cosmological perturbations on the brane. The power spectrum is scale invariant for large $k$'s and scales as $k$ for small $k$'s.

We  then tested this prediction for power spectrum against the cosmological observations.  The transition is characterized by a decay of $1/k$ for scales where the power spectrum is highly constrained by data \cite{Ade:2015lrj} as shown in Fig. \ref{fig.plots} (right). We found that our model is broadly consistent with non-parametric reconstruction of primordial power spectrum, but is disfavoured compared to a pure power-law at 2.7$\sigma$ level. We finally
 outlined various theoretical constraints on the nucleation temperature and 5D Planck mass in the HBB model, and found that  the best fit nucleation temperature of the 3-brane was  at least  3 orders of magnitude larger than the 5D Planck mass.

This first attempt to understand the detailed consequences of the HBB model for cosmology relies on several simplifying assumptions that can be relaxed in future work. Some of the issues that remain to be tackled are:

\begin{enumerate}

\item Perhaps our most perplexing finding was that our best-fit model violated the holographic entropy bound by 8 orders of magnitude. It is not yet clear whether this is a feature or a bug!

\item  It would be interesting to study how brane cosmological perturbations will be affected by the large-scale curvature of the bulk (in a 5D Schwarzschild of Kerr spacetime). 

\item Other observables that remain to be computed are the amplitude of tensor modes and the non-gaussianity, although we do not expect them to be significant.  
 
\item Given that the speed of sound for a relativistic 5D atmosphere is $c_s = c/2$, one expect ${\cal O}(0.2)$ relativistic corrections to the atmosphere profile, which we have ignored. This could  affect the functional shape of the power spectrum at a similar level, potentially improving (or worsening) the fit to the data.   A related issue is whether the hydrostatic equilibrium profile for the relativistic  thin atmosphere is stable.
\end{enumerate}

To conclude, while we believe the 5D holographic big bang remains an intriguing possibility for the origin of our universe, there remain empirical and theoretical challenges to its status amongst various scenarios for the early universe cosmology that should be addressed in future work.

\section*{Acknowledgements}
This work has been partially supported by the National Science and Engineering Research Council (NSERC), University of Waterloo, and Perimeter Institute for Theoretical Physics (PI). Research at PI is supported by the 
Government of Canada through the Department of Innovation, Science and Economic Development Canada and by the Province of Ontario through 
the Ministry of Research, Innovation and Science.

\bibliography{Databaze}

\providecommand{\href}[2]{#2}\begingroup\raggedright\begin{thebibliography}{10}

\bibitem{Plankresults2015}
{\bf Planck} Collaboration, P.~A.~R. Ade {\em et.~al.}, {\it {Planck 2015
  results. XIII. Cosmological parameters}},
  \href{http://xxx.lanl.gov/abs/1502.0158}{{\tt arXiv:1502.0158}}.

\bibitem{Holobigbang2013}
R.~Pourhasan, N.~Afshordi, and R.~B. Mann, {\it {Out of the White Hole: A
  Holographic Origin for the Big Bang}},  {\em JCAP} {\bf 1404} (2014) 005,
  [\href{http://xxx.lanl.gov/abs/1309.1487}{{\tt arXiv:1309.1487}}].

\bibitem{Dvali:2000hr}
G.~R. Dvali, G.~Gabadadze, and M.~Porrati, {\it {4-D gravity on a brane in 5-D
  Minkowski space}},  {\em Phys. Lett.} {\bf B485} (2000) 208--214,
  [\href{http://xxx.lanl.gov/abs/hep-th/0005016}{{\tt hep-th/0005016}}].

\bibitem{Gregory:2007xy}
R.~Gregory, N.~Kaloper, R.~C. Myers, and A.~Padilla, {\it {A New perspective on
  DGP gravity}},  {\em JHEP} {\bf 10} (2007) 069,
  [\href{http://xxx.lanl.gov/abs/0707.2666}{{\tt arXiv:0707.2666}}].

\bibitem{Mathur:2008kg}
S.~D. Mathur, {\it {Tunneling into fuzzball states}},  {\em Gen. Rel. Grav.}
  {\bf 42} (2010) 113--118, [\href{http://xxx.lanl.gov/abs/0805.3716}{{\tt
  arXiv:0805.3716}}].

\bibitem{Witten1982481}
E.~Witten, {\it Instability of the kaluza-klein vacuum},  {\em Nuclear Physics
  B} {\bf 195} (1982), no.~3 481 -- 492.

\bibitem{Lunin:2002qf}
O.~Lunin and S.~D. Mathur, {\it {Statistical interpretation of Bekenstein
  entropy for systems with a stretched horizon}},  {\em Phys. Rev. Lett.} {\bf
  88} (2002) 211303, [\href{http://xxx.lanl.gov/abs/hep-th/0202072}{{\tt
  hep-th/0202072}}].

\bibitem{Pospelov:2010hj}
M.~Pospelov and J.~Pradler, {\it {Big Bang Nucleosynthesis as a Probe of New
  Physics}},  {\em Ann. Rev. Nucl. Part. Sci.} {\bf 60} (2010) 539--568,
  [\href{http://xxx.lanl.gov/abs/1011.1054}{{\tt arXiv:1011.1054}}].

\bibitem{Azizi:2011ys}
T.~Azizi, M.~Sadegh~Movahed, and K.~Nozari, {\it {Observational Constraints on
  the Normal Branch of a Warped DGP Cosmology}},  {\em New Astron.} {\bf 17}
  (2012) 424--432, [\href{http://xxx.lanl.gov/abs/1111.3195}{{\tt
  arXiv:1111.3195}}].

\bibitem{Ade:2015rim}
{\bf Planck} Collaboration, P.~A.~R. Ade {\em et.~al.}, {\it {Planck 2015
  results. XIV. Dark energy and modified gravity}},
  \href{http://xxx.lanl.gov/abs/1502.0159}{{\tt arXiv:1502.0159}}.

\bibitem{1996ApJ...469..437S}
U.~{Seljak} and M.~{Zaldarriaga}, {\it {A Line-of-Sight Integration Approach to
  Cosmic Microwave Background Anisotropies}},  {\em The Astrophysical Journal}
  {\bf 469} (Oct., 1996) 437,
  [\href{http://xxx.lanl.gov/abs/astro-ph/9603033}{{\tt astro-ph/9603033}}].

\bibitem{1998ApJ...494..491Z}
M.~{Zaldarriaga}, U.~{Seljak}, and E.~{Bertschinger}, {\it {Integral Solution
  for the Microwave Background Anisotropies in Nonflat Universes}},  {\em The
  Astrophysical Journal} {\bf 494} (Feb., 1998) 491--502,
  [\href{http://xxx.lanl.gov/abs/astro-ph/9704265}{{\tt astro-ph/9704265}}].

\bibitem{2002PhRvD..66j3511L}
A.~{Lewis} and S.~{Bridle}, {\it {Cosmological parameters from CMB and other
  data: A Monte Carlo approach}},  {\em Phys. Rev. D} {\bf 66} (Nov., 2002)
  103511, [\href{http://xxx.lanl.gov/abs/astro-ph/0205436}{{\tt
  astro-ph/0205436}}].

\bibitem{2000ApJ...538..473L}
A.~{Lewis}, A.~{Challinor}, and A.~{Lasenby}, {\it {Efficient Computation of
  Cosmic Microwave Background Anisotropies in Closed Friedmann-Robertson-Walker
  Models}},  {\em The Astrophysical Journal} {\bf 538} (Aug., 2000) 473--476,
  [\href{http://xxx.lanl.gov/abs/astro-ph/9911177}{{\tt astro-ph/9911177}}].

\bibitem{2012JCAP...04..027H}
C.~{Howlett}, A.~{Lewis}, A.~{Hall}, and A.~{Challinor}, {\it {CMB power
  spectrum parameter degeneracies in the era of precision cosmology}},  {\em
  JCAP} {\bf 4} (Apr., 2012) 027,
  [\href{http://xxx.lanl.gov/abs/1201.3654}{{\tt arXiv:1201.3654}}].

\bibitem{2013PhRvD..87j3529L}
A.~{Lewis}, {\it {Efficient sampling of fast and slow cosmological
  parameters}},  {\em Phys. Rev. D} {\bf 87} (May, 2013) 103529,
  [\href{http://xxx.lanl.gov/abs/1304.4473}{{\tt arXiv:1304.4473}}].

\bibitem{campnotes}
A.~{Lewis}, {\it {CAMB Notes}},  {\em http://cosmologist.info/notes/CAMB.pdf}.

\bibitem{2011MNRAS.416.3017B}
F.~{Beutler}, C.~{Blake}, M.~{Colless}, D.~H. {Jones}, L.~{Staveley-Smith},
  L.~{Campbell}, Q.~{Parker}, W.~{Saunders}, and F.~{Watson}, {\it {The 6dF
  Galaxy Survey: baryon acoustic oscillations and the local Hubble constant}},
  {\em mnras} {\bf 416} (Oct., 2011) 3017--3032,
  [\href{http://xxx.lanl.gov/abs/1106.3366}{{\tt arXiv:1106.3366}}].

\bibitem{2011MNRAS.418.1707B}
C.~{Blake} and others., {\it {The WiggleZ Dark Energy Survey: mapping the
  distance-redshift relation with baryon acoustic oscillations}},  {\em mnras}
  {\bf 418} (Dec., 2011) 1707--1724,
  [\href{http://xxx.lanl.gov/abs/1108.2635}{{\tt arXiv:1108.2635}}].

\bibitem{2012MNRAS.427.3435A}
L.~{Anderson} {\em et.~al.}, {\it {The clustering of galaxies in the SDSS-III
  Baryon Oscillation Spectroscopic Survey: baryon acoustic oscillations in the
  Data Release 9 spectroscopic galaxy sample}},  {\em mnras} {\bf 427} (Dec.,
  2012) 3435--3467, [\href{http://xxx.lanl.gov/abs/1203.6594}{{\tt
  arXiv:1203.6594}}].

\bibitem{2012MNRAS.423.3430B}
F.~{Beutler}, C.~{Blake}, M.~{Colless}, D.~H. {Jones}, L.~{Staveley-Smith},
  G.~B. {Poole}, L.~{Campbell}, Q.~{Parker}, W.~{Saunders}, and F.~{Watson},
  {\it {The 6dF Galaxy Survey: z 0 measurements of the growth rate and
  {$\sigma$}$_{8}$}},  {\em mnras} {\bf 423} (July, 2012) 3430--3444,
  [\href{http://xxx.lanl.gov/abs/1204.4725}{{\tt arXiv:1204.4725}}].

\bibitem{2012MNRAS.427.2132P}
N.~{Padmanabhan}, X.~{Xu}, D.~J. {Eisenstein}, R.~{Scalzo}, A.~J. {Cuesta},
  K.~T. {Mehta}, and E.~{Kazin}, {\it {A 2 per cent distance to z = 0.35 by
  reconstructing baryon acoustic oscillations - I. Methods and application to
  the Sloan Digital Sky Survey}},  {\em mnras} {\bf 427} (Dec., 2012)
  2132--2145, [\href{http://xxx.lanl.gov/abs/1202.0090}{{\tt
  arXiv:1202.0090}}].

\bibitem{2014MNRAS.441...24A}
L.~{Anderson} and others., {\it {The clustering of galaxies in the SDSS-III
  Baryon Oscillation Spectroscopic Survey: baryon acoustic oscillations in the
  Data Releases 10 and 11 Galaxy samples}},  {\em mnras} {\bf 441} (June, 2014)
  24--62, [\href{http://xxx.lanl.gov/abs/1312.4877}{{\tt arXiv:1312.4877}}].

\bibitem{2014MNRAS.439.3504S}
L.~{Samushia} and others., {\it {The clustering of galaxies in the SDSS-III
  Baryon Oscillation Spectroscopic Survey: measuring growth rate and geometry
  with anisotropic clustering}},  {\em mnras} {\bf 439} (Apr., 2014)
  3504--3519, [\href{http://xxx.lanl.gov/abs/1312.4899}{{\tt
  arXiv:1312.4899}}].

\bibitem{2015MNRAS.449..835R}
A.~J. {Ross}, L.~{Samushia}, C.~{Howlett}, W.~J. {Percival}, A.~{Burden}, and
  M.~{Manera}, {\it {The clustering of the SDSS DR7 main Galaxy sample - I. A 4
  per cent distance measure at z = 0.15}},  {\em mnras} {\bf 449} (May, 2015)
  835--847, [\href{http://xxx.lanl.gov/abs/1409.3242}{{\tt arXiv:1409.3242}}].

\bibitem{Ade:2015zua}
{\bf Planck} Collaboration, P.~A.~R. Ade {\em et.~al.}, {\it {Planck 2015
  results. XV. Gravitational lensing}},  {\em Astron. Astrophys.} {\bf 594}
  (2016) A15, [\href{http://xxx.lanl.gov/abs/1502.0159}{{\tt
  arXiv:1502.0159}}].

\bibitem{Aghanim:2015xee}
{\bf Planck} Collaboration, N.~Aghanim {\em et.~al.}, {\it {Planck 2015
  results. XI. CMB power spectra, likelihoods, and robustness of parameters}},
  {\em Astron. Astrophys.} (2015)
  [\href{http://xxx.lanl.gov/abs/1507.0270}{{\tt arXiv:1507.0270}}].

\bibitem{Ade:2015fva}
{\bf Planck} Collaboration, P.~A.~R. Ade {\em et.~al.}, {\it {Planck 2015
  results. XXIV. Cosmology from Sunyaev-Zeldovich cluster counts}},  {\em
  Astron. Astrophys.} {\bf 594} (2016) A24,
  [\href{http://xxx.lanl.gov/abs/1502.0159}{{\tt arXiv:1502.0159}}].

\bibitem{2012ApJ...755...70R}
C.~L. {Reichardt} {\em et.~al.}, {\it {A Measurement of Secondary Cosmic
  Microwave Background Anisotropies with Two Years of South Pole Telescope
  Observations}},  {\em apj} {\bf 755} (Aug., 2012) 70,
  [\href{http://xxx.lanl.gov/abs/1111.0932}{{\tt arXiv:1111.0932}}].

\bibitem{2014JCAP...04..014D}
S.~{Das} {\em et.~al.}, {\it {The Atacama Cosmology Telescope: temperature and
  gravitational lensing power spectrum measurements from three seasons of
  data}},  {\em jcap} {\bf 4} (Apr., 2014) 014,
  [\href{http://xxx.lanl.gov/abs/1301.1037}{{\tt arXiv:1301.1037}}].

\bibitem{Ade:2015lrj}
{\bf Planck} Collaboration, P.~A.~R. Ade {\em et.~al.}, {\it {Planck 2015
  results. XX. Constraints on inflation}},
  \href{http://xxx.lanl.gov/abs/1502.0211}{{\tt arXiv:1502.0211}}.

\bibitem{1990eaun.book.....K}
E.~W. {Kolb} and M.~S. {Turner}, {\em {The early universe.}}
\newblock 1990.

\end{thebibliography}\endgroup


\providecommand{\href}[2]{#2}\begingroup\raggedright\endgroup
\bibliographystyle{JHEP}

\appendix

\section{Inhomogeneous Cosmological perturbations on the bulk}\label{app:perturbations}

If we consider the metric \eqref{bulkmetric} and homogeneous perturbations of the form \eqref{homopert} the induced metric on the brane is 
$\gamma_{\mu\nu}=g_{\alpha\beta}e^{\alpha}_{\mu}e^{\beta}_{\nu}$ that to first order in $\epsilon$ reads: 
\beq
ds^2_{\text{brane}}=[(f_0'^2-1)-2 \epsilon\left(\text{$\hat{\hat{\Psi}}_5^0 $}(f) f_0'^2+\text{$ \hat{\Phi}_5^0 $}(f)\right)]dt^2+
[1-2 \epsilon  \text{$ \hat{\Psi}_5^0 $}(f)][dx^2+dy^2+dz^2]\,,
\label{branemetric}
\eeq
where  $f'=\frac{\text{d}f}{\text{d}t}$. The induced metric on the brane has to be able to describe a Friedmann universe for which we make the following identifications
\beq
-d\tau^2=[(f_0'^2-1)-2 \epsilon \left(\hat{\Psi}_5^0  f_0'^2+ \hat{\Phi}_5^0 \right)]dt^2\,, \quad a^2=[1-2  \epsilon \hat{\Psi}_5^0]\,,
\eeq
where $\tau$ is the proper time of the brane and $a$ is the scale factor. On top of the homogeneous perturbations we are now going to consider anisotropies
\beq
\Phi_5=\epsilon\Phi_5^0(w)+\epsilon_1 \Phi_5^1(x^{\alpha})\,, \quad \Psi_5=\epsilon\Psi_5^0(w)+\epsilon_1 \Psi_5^1(x^{\alpha})\,, \quad f=f_0(t)+\epsilon_1 f_1(x^{\alpha})\,,
\eeq
where $\epsilon_1\ll1$ is a parameter that controls the anisotropic perturbations and $\epsilon \epsilon_1 \ll \epsilon, \epsilon_1$. The induced metric  can be written as
\beq
ds^2_{\text{brane}}=-[1+\frac{2\epsilon_1}{f_0'^2-1}(\hat{\Phi}_5^1+\hat{\Psi}_5^1f_0^2-f_0'f_1')]d\tau^2+2a\epsilon_1
\frac{f_0'f_{1,i}}{\sqrt{f_0'^2-1}}dx^id\tau+a^2(1-2\hat{\Psi}_5^1\epsilon_1)dx^2\,,
\label{induced4d}
\eeq
 where $\hat{\Psi}_5=\Psi_5(w=f(x^\mu))$ are the metric functions projected to the brane. The general form of a 4D metric including scalar  and vector  cosmological perturbation in 4D are that contains all the terms in Eq. \eqref{induced4d} 
\beq
ds^2_{\text{brane}}=-(1+2 \phi_4)d\tau^2-2a B_id\tau dx^i+a^2(1-2\psi_4)dx^2\,,
\eeq
and thus we identify
\beq
\phi_4=\frac{\epsilon_1}{f_0'^2-1}(\hat{\Phi}_5^1+\hat{\Psi}_5^1f_0^2-f_0'f_1')\,, \quad \psi_4=\hat{\Psi}_5^1\epsilon_1\,, \quad B_i=\epsilon_1\frac{f_0'f_{1,i}}{\sqrt{f_0'^2-1}}\,.
\eeq
Note that the Newtonian gauge on the bulk does not translate into a Newtonian gauge on the brane and that some perturbations that are scalars in 5D are projected as a vectorial perturbation component in 4D. The scalar gauge invariant quantities can be constructed from Eq.\eqref{induced4d} as 
\bea
\Phi_4&=&\phi_4-\partial_{\tau}(aB)\,,\\
\Psi_4&=&\psi_4+HaB\,, \label{4metper1}
\eea
where, $H$ is the Hubble constant and $B$ is the scalar part of the vector metric perturbation $B_i$. From our construction the Hubble constant is 
\beq
H=\frac{\dot{a}}{a}=-\frac{\epsilon \hat{\Psi}_5^0}{a^2}\,,
\label{perturbed_hubble}
\eeq
and thus Eq.\eqref{4metper1} reduces to $\Psi_4=\psi_4$ to first order in $\epsilon, \epsilon_1$.

\section{Derivation of the power spectrum} \label{app:power}

With use of the Poisson equation in 4D
\beq
\nabla^2\Psi_5(y)=\frac{8\pi G_5}{3}\rho_5(y)\,,
\label{poisson5A}
\eeq
and the expression for the energy density correlation function 
\beq
  \submin{\rho_5(y_1)\rho_5(y_2)}\simeq
  \alpha (T_5)^6 \delta^4(y_1-y_2)\,,
  \eeq
 we can write the 2-point correlation function of $\Psi_5$ using the Green's function for the
Laplacian operator
\beq
\langle \Psi_5(x_1)\Psi_5(x_2) \rangle=\alpha \bigg(\frac{8\pi G_5}{3}\bigg)^2\bigg(\frac{1}{4\pi^2}\bigg)^2\int d^4y \frac{(T_5(y))^6}{|y-x_1|^2|y-x_2|^2}\,.
\eeq
The junction condition between the 5D and 4D metrics \eqref{bulkmetric} and \eqref{branemetric} allow us to compute $\langle\Psi_4({\bf{x}_1})\Psi_4({\bf{x}_2})\rangle$
\begin{eqnarray}
\langle\Psi_4({\bf{x}_1})\Psi_4({\bf{x}_2})\rangle&=& \langle\Psi_5({\bf{x}_1},x_{1w}=0)\Psi_4({\bf{x}_2},x_{2w}=0)\rangle\nonumber \\
&=&\alpha \bigg(\frac{2 G_5}{3\pi}\bigg)^2\!\!\!
\int d^3{\bf{y}_3}dy_w\frac{(T_5(y_w))^6}{(|{\bf{x}_1}-{\bf{y}_3}|^2+|y_w|^2)(|{\bf{x}_2}-{\bf{y}_3}|^2+|y_w|^2)} 
\label{2point}
\end{eqnarray}
where we have decomposed the bulk coordinate as $y=({\bf{y}_3},y_w)$ and we have set the temperature to be just a function of the $w$ direction of the bulk. 
Combining expressions \eqref{cur_pert}, \eqref{powerspectrum}, \eqref{2point} and setting ${\bf{x}_1}=0$ we have 
\begin{eqnarray}
 P_{\zeta}(k)&=&\alpha \bigg(\frac{2 G_5}{3\pi \,\Delta \text{ln}H}\bigg)^2\int \frac{d^3{\bf{y}_3}\,dy_w \,(T_5(y_w))^6}{|{\bf{y}_3}|^2+|y_w|^2}\int d^3x \frac{e^{i {\bf{k\cdot x}}}}{|{\bf{x}}-{\bf{y}_3}|^2+|y_w|^2}\,, \nonumber \\
 &=&\alpha \bigg(\frac{2 G_5}{3\pi  \,\Delta \text{ln}H}\bigg)^2\int \frac{d^3{\bf{y}_3}\,dy_w \,(T_5(y_w))^6}{|{\bf{y}_3}|^2+|y_w|^2}\int d^3{\bf{x}}' \frac{e^{i {\bf{k\cdot x'}}}e^{i {\bf{k\cdot {\bf{y}_3}}}}}{|{\bf{x}}'|^2+|y_w|^2}\,,  \nonumber \\
 &=&\alpha \bigg(\frac{2 G_5}{3\pi  \,\Delta \text{ln}H}\bigg)^2\int dy_w\, (T_5(y_w))^6\bigg (\int d^3{\bf{x}}' \frac{e^{i {\bf{k\cdot x'}}}}{|{\bf{x}}'|^2+|y_w|^2}\bigg)^2\,,  \nonumber \\
 &=&\alpha \bigg(\frac{2 G_5}{3\pi \,\Delta \text{ln}H}\bigg)^2\frac{1}{(4\pi k)^2}\int dy_w\, e^{-2|y_w|k}\,(T_5(y_w))^6\,,  \label{power_spectrum}
 \end{eqnarray}
where we have performed the coordinate transformation ${\bf{x}}'={\bf{x}}-{\bf{y}_3}$, and use the result 
$\int\frac{ d^3x\, e^{i{\bf{k\cdot x}}}}{|x|^2+|m|^2}=\frac{e^{-k|m|}}{4\pi k}$.

With this we are able to write
\beq
{\cal{P}}(k)=\frac{k^3}{2\pi^2}P_{\zeta}(k)=\beta\, k \int_0^{\infty} dw\, e^{-2|w|k}\,(T_5(w))^6\,,
\label{normalizedpsA}
\eeq
where $\beta=\frac{\alpha}{2}\bigg(\!\frac{G_5}{6\Delta \text{ln}H \,\pi^3}\!\bigg)^2$. Note that we are in
integrating between $[0,\infty)$ because of the $\mathbb{Z}_2$ symmetry. Because the temperature profile is analytic it admits a Taylor expansion of the form
 \beq
 (T_5(w))^6 =\sum_{n=0}^{\infty}\frac{T^{6^{(n)}}(0,\bar{\gamma},\bar{w})}{n!}w^n\,,
 \label{taylor}
 \eeq
  and thus the power spectrum Eq.\eqref{normalizedpsA} admits a series decomposition of the form
  \beq
{\cal{P}}(k)=\beta\sum_{n=0}^{\infty}\frac{T^{6^{(n)}}(0,\bar{\gamma},\bar{w})}{k^{n}}2^{-1-n}\,.
\label{asympower}
\eeq
This last expression implies that for large $k$ the correction to a scale invariant power spectrum goes as $1/k$. If the integral in Eq.\eqref{normalizedpsA} is performed in the rage $(-\infty,\infty)$ the the power spectrum series would be 
\beq
{\cal{P}}(k)=\beta\sum_{n=0}^{\infty}\frac{T^{6^{(2n)}}(0,\bar{\gamma},\bar{w})}{k^{2n}}2^{-1-2n}\,,
\label{sympower}
\eeq
giving a correction from scale invariant that goes as $1/k^2$ for large $k$. This correction renders model  disfavourable in comparison with the symmetric case, where we have a $1/k$ decay. This is the reason why we work with the symmetric integral Eq.\eqref{normalizedpsA}. 

\end{document}